\documentclass[reprint, aps,amssymb,amsmath,showpacs] {revtex4-2}%

\usepackage{graphicx}
\usepackage{dcolumn}
\usepackage{bm}
\usepackage{siunitx}%
\usepackage[version=4]{mhchem}
\usepackage{xspace}
\usepackage{booktabs}
\usepackage{float}
\usepackage{placeins}
\usepackage{lineno}
\usepackage{subfigure}
\usepackage{enumerate}
\usepackage{hyperref}
\usepackage{enumitem}

\newcommand*{\pT}{\ensuremath{p_\mathrm{T}}\xspace}
\newcommand*{\pTreco}{\ensuremath{p_\mathrm{T,jet}^\mathrm{reco}}\xspace}
\newcommand*{\pThat}{\ensuremath{\hat{p}_\mathrm{T}}\xspace}
\newcommand*{\Ajet}{\ensuremath{A_\mathrm{jet}}\xspace}
\newcommand*{\pTtruth}{\ensuremath{p_\mathrm{T,jet}^\mathrm{truth}}\xspace}
\newcommand*{\pTcorr}{\ensuremath{p_\mathrm{T,jet}^\mathrm{corr}}\xspace}
\newcommand*{\residpt}{\ensuremath{\delta p_\mathrm{T,jet}}\xspace}
\newcommand*{\pTresid}{\ensuremath{\delta p_\mathrm{T,jet}}\xspace}
\newcommand*{\pTjet}{\ensuremath{p_\mathrm{T,jet}}\xspace}
\newcommand*{\rhobkg}{\ensuremath{\rho_\mathrm{bkg}}\xspace}
\newcommand*{\kT}{\ensuremath{k_\mathrm{T}}\xspace}
\newcommand*{\gevc}[1]{\ensuremath{\SI{#1}{GeV/\mathit{c}}}\xspace}
\newcommand*{\NN}[1]{\ensuremath{\mathrm{NN}_\mathrm{#1}\xspace}}
\newcommand*{\antikT}{\ensuremath{\mathrm{anti\text{-}}k_\mathrm{T}}\xspace}
\newcommand*{\sNNcc}{\ensuremath{\sqrt{s_\mathrm{NN}}=\SI{200}{GeV}}\xspace}

\newcommand*{\RAA}{\ensuremath{R_\mathrm{AA}}\xspace}
\newcommand*{\RAALeadJet}{\ensuremath{R_\mathrm{AA}^\mathrm{LeadJet}}\xspace}

\newcommand*{\Mct}{\ensuremath{\mathcal{M}\left(\pTtruth,\pTreco\right)}\xspace}

\begin{document}
\preprint{APS/123-QED} 
\title{
Neural network biased corrections:\\ Cautionary study in background corrections for quenched jets
}

\author{D.~Stewart}\affiliation{Wayne State University, Detroit, Michigan 48201}
\author{J.~Putschke}\affiliation{Wayne State University, Detroit, Michigan 48201}

 

\date{\today}

\begin{abstract} 

\noindent 

Jets clustered from heavy ion collision measurements combine a dense background of particles with those actually resulting from a hard partonic scattering. The background contribution to jet transverse momentum (\pT) may be corrected by subtracting the collision average background; however, the background inhomogeneity limits the resolution of this correction. Many recent studies have embedded jets into heavy ion backgrounds and demonstrated a markedly improved background correction is achievable by using neural networks (NNs) trained with aspects of jet substructure which are used to map measured jet \pT to the embedded truth jet \pT. However, jet quenching in heavy ion collisions modifies jet substructure, and correspondingly biases the NNs' background corrections. This study investigates those biases by using simulations of jet quenching in central Au+Au collisions at \sNNcc with hydrodynamically modeled quark-gluon plasma (QGP) evolution. To demonstrate the magnitude of the effect of such biases in measurement, a leading jet nuclear modification factor (\RAA) is calculated and reported using the NN background correction on jets quenched utilizing a brick of QGP.

\end{abstract}
\maketitle

\section{Introduction}\label{sec:Introduction}

Heavy ion collisions at ultra-relativistic speeds were first proposed to provide the experimental conditions necessary to create and study the quark-gluon plasma (QGP), a novel phase of matter predicted by the completion of the standard model of physics in which quarks and gluons are deconfined throughout an extended volume \cite{Harris:1996zx}. Correspondingly, QGP signals were observed with the first data from collisions at the Relativistic Heavy Ion Collider (RHIC) \cite{BRAHMS:2004adc,PHOBOS:2004zne,STAR:2005gfr,PHENIX:2004vcz} and then the Large Hadron Collider (LHC) \cite{ATLAS:2010isq,CMS:2011iwn,ALICE:2010suc}. The study of QGP formation, properties, evolution, and hadronization, has remained the central motivation of ultra-relativistic heavy ion collision research for the past quarter century. The reader is referred to \cite{Niida:2021wut} for a recent overview.

Rare, high-$Q^{2}$ scattered partons, jets, in ultra-relativistic heavy-ion collisions occur early in the collision evolution. Therefore, they provide intrinsic colored probes which undergo scattering and induced gluon emmission, i.e., jet quenching, while traversing the QGP, which is a hot, dense, colored medium. The simplest experimental proxy for these initial hard partons are high transverse momentum (\pT) hadrons, and indeed one of the first strong evidences for QGP formation, predicted in 1982 by J. D. Bjorken (see \cite{Baier:2002tc}), was the disappearance (quenching) of high-\pT hadrons in central (i.e., head-on) Au+Au collisions at RHIC in 2004 \cite{STAR:2004amg}. 

In subsequent years, infrared-safe and collinear-safe clustering algorithms \cite{CTEQ:1993hwr} have been used to combine clusters of collimated particles into aggregate objects called jets. Jets provide many advantages over individual hadrons, not least of which is they provide a better proxy for the initial scattered parton with more of the originating \pT recaptured experimentally. However, they are also complex objects collecting all measured particles within a physical geometric acceptance, generally measured in coordinates of pseudo-rapidity ($\eta$) and azimuth ($\phi$). In heavy ion collisions this collection includes many ``background particles'': particles not originating from the initiating partonic hard-scattering, but rather from many of the other participating colliding nuclei or even additional (soft) scatterings of the originating hard-scattered partons. For a recent overview of jet measurements see \cite{Cunqueiro:2021wls}.

At RHIC, jets have been measured up to around \SI{40}{GeV/\mathit{c}} \cite{PHENIX:2024hkz,STAR:2023hye,STAR:2024nwm}. For comparison, in the 5\% most central collisions of Au+Au ions at \sNNcc, the background particles contribute, on average, around \SI{45}{GeV/\mathit{c}} of additional \pT to \antikT jets with $R=0.4$, where $R$ is the jet resolution parameter and is approximately equal to the jet radius in $\eta\times\phi$ space \cite{Cacciari:2008gp}.

The area-based (AB) method is a common experimental procedure to correct for the heavy background. Proposed in 2007 \cite{Cacciari:2007fd}, and later implemented by the authors in the FastJet package \cite{Cacciari:2011ma}, this method measures the median background density (\rhobkg) in each event and then corrects the \pT of measured jets (\pTreco) by \rhobkg scaled by the jet area: $\pTcorr\equiv\pTreco-\rhobkg\Ajet$. The jet areas (\Ajet) for \antikT jets are relatively stable \cite{Cacciari:2008gp} and therefore have little impact on the AB correction, and the effect of the leading \pT jets on \rhobkg is algorithmically minimized by disregarding the one or two highest \pT jets while calculating \rhobkg. As a result, the AB method's corrections utilize principally information from the event background, and is therefore insensitive to jet substructure.

The resolution of the AB method, parametrized in this paper by the distribution of the residual errors between the corrected jet \pT and the actual truth jet \pT ($\pTresid\equiv\pTcorr-\pTtruth$), is dominated by region-to-region fluctuations in the background densities. In experiment, these must be corrected statistically, typically at the same time as the jet energy spectra and resolution due to detector effects. These are on the order of \SI{8}{GeV/\mathrm{c}} per $R=0.4$ jet in central Au+Au collisions at RHIC and are the principle limitation on jet measurement resolution in these heavy-ion collisions.  Alternatively, at kinematics achievable in collisions at the LHC, jets are available at much higher values of \pT, but the lower-boundary of \pT available to jet measurements is also limited by the resolution of the background fluctuations.

In order to improve the resolution of jet measurements beyond the fundamental limitation imposed by background fluctuations, additional information is needed. More than 40 years of development incorporating both perturbative calculations and models for fragmentation and hadronization have yielded well-tuned Monte Carlo simulators for high-energy $pp$ collisions \cite{SJOSTRAND2020106910,Bierlich:2022pfr,Bellm:2015jjp}. It is natural to take the jet parameters of these $pp$ simulations and see if some combination of observable signals from jet substructure can be combined with the \rhobkg, \pTreco, and \Ajet of the AB method, to improve the jet \pT  background subtraction in heavy-ion collisions. This search is facilitated by the powerful tools afforded by Machine Learning (ML), which can learn and apply correlations of essentially arbitrary dimensionality. Indeed, it has been demonstrated in prior publications (and reconfirmed in this present paper) that neural networks (NNs) gain significant discriminatory power in correcting jets for background fluctuations if they are provided the \pT values of the leading jet constituents or even just the total number of jet constituents  \cite{Haake:2018hqn,ALICE:2023waz,Mengel:2024fcl}.

A principle motivation to study jet \pT in heavy ion collisions is to measure quenching, in which the jet substructure is modified through partonic energy interactions with the QGP. Consequently, there is significant uncertainty about substructure in quenched jets relative to that in vacuum jets. This relative uncertainty propagates in any calculation dependent on jet substructure, e.g., specific detector response effects, and is also present in selections of rare-$pp$ collisions, such as those resulting in ultra-high particle multiplicities. The difficulties posed by using jet substructure to correct for heavy backgrounds to measure jet quenching are particularly problematic: the partonic energy loss modifies the jet substructure, and a particular jet substructure must be assumed for the background correction to measure quenching. Inaccuracy in the assumed substructure introduces a bias in the background \pT correction.
This paper investigates those biases by using NNs trained on $pp$ jets and using them to background correct quenched jets. In order to indicate how such a bias may propagate in actual jet measurements, it also reports an error of a simulated measurement of the nuclear modification factor \RAA of a quenched jet spectrum. 

A measurement of \RAA using both area-based and ML-based background corrections been published for collisions at LHC in \cite{ALICE:2023waz} in which the ML-based measurements have significantly higher systematic errors but reach lower jet \pT values than previously accessible using the area-based method. This current paper is complementary in the sense that it investigates the effects in collisions at RHIC energies, with its associated steeper jet \pT spectrum and different background conditions.

This study utilizes the JETSCAPE framework \cite{Putschke:2019yrg} to generate jets in both $pp$ collisions and central Au+Au collisions at \sNNcc, using tune parameters from \cite{JETSCAPE:2020mzn,JETSCAPE:2022jer}. The Au+Au collision simulations hydrodynamically model the QGP production and evolution and are hereafter referred to as ``hydro events'' in this study. This collision system and energy has been selected to match the collisions scheduled for RHIC's 2025 run. The simulations in these hydro events is the best available to use with jet quenching Monte Carlos (MCs), and this is the first study to use them to investigate ML. They do not, however, model medium response to the jets. The hydro events also provide realistic hadron backgrounds for central Au+Au events. In all cases, for the $pp$ and heavy ion events, the particles are used as-provided by the Monte-Carlo, without any additional simulation to account for detector effects or efficiencies.

Using the jets from the $pp$ events along with the backgrounds from the hydro events, we train NNs and reconfirm the findings from previous studies (e.g., \cite{Haake:2018hqn,Mengel:2024fcl}) which show the ability of NNs to \pT correct jets for heavy backgrounds. We do this by embedding $pp$ jets into heavy ion backgrounds  -- using both charged and neutral particles -- from the hydro events, clustering the combined events, and geometrically matching the resulting jets to the $pp$ jets clustered in vacuum. We train NNs to map the matched composite jets to the vacuum jets using various selections of jet parameters. The goodness of the resulting correction is defined by the distributions of the \pT residual errors, \residpt.

While they are the best option for jet quenching simulations, the hydro events are also very computationally expensive. In order to generate a full spectrum of quenched jets, we also used jets simulated by quenching in fixed-length ``bricks'' of QGP. These are much faster to generate and allowed the generation of the several million jets which compose the full spectrum. These were generated in order to provide a standard jet quenching measurement, an \RAA, in order to discuss the biases.

We report the jet quenching via the modification of the jet fragmentation in the hydro and in brick events using brick lengths up to six fm. We also report the \pTresid distributions for the NN background corrections for these jets. To do this, the brick jets were embedded into backgrounds from the hydro events. This creates reasonable, computationally-cheap, approximation of a hydro event; however, it also destroys effects from variable path lengths and the evolving medium on jet quenching. These \pTresid distribution studies serve two purposes. First, they measure the progression of biases in \pTresid with increasing amounts of quenching. Second, they determine the brick length whose quenching is most equivalent to the quenching in hydro events. That brick length is then used for a full-spectrum of quenched jets.

Finally, to demonstrate the magnitude of the effect on final jet measurements that can result from the propagation of the \pTresid biases,  we simulate an \RAA measurement for the leading, i.e. highest-\pT, jet per event. In this simulation, the ``data'' consists of the full spectrum of jets quenched in QGP bricks and embedded into hydro backgrounds. The measured jets are background corrected using the NNs, which were trained on $pp$ jets embedded in the hydro backgrounds. Using the same algorithm as used in experimental measurements, the process inefficiencies are corrected on an ensemble level to a final ``measured'' \pT spectrum. The ratio of this spectrum to the vacuum $pp$ spectrum (\RAALeadJet) is presented and compared to the actual \RAALeadJet from the brick events. 

\section{Data Simulation: Production and Processing}

\noindent\textit{\footnotesize Note: The libraries used are listed in Appendix~\ref{app:software}. The code and notebooks used  are archived online at \hyperlink{https://github.com/david-stewart/jet_and_thermal}{https://github.com/david\-stewart/jet\_and\_thermal}.}

\subsection{Jet Simulation}

The JETSCAPE framework was used to simulate hard scatterings of initial partons, along with their subsequent evolution and hadronization. Jets were generated using the input Monte Carlo input parameter \pThat, which constrains the transverse momentum of the initiating partons (IP) in the hard scattering. (For illustration, refer to Fig.~\ref{fig:pTtruthPerIP} in the appendix, which shows the \pT distributions of jets resulting from different \pThat selections.) We generated groups of jets at both discreet \pThat values -- which correspond to tightly clustered jet \pT -- and from groups of events with a range of \pThat's, each weighted to account for the \pThat cross sections, in order to simulate continuous, steeply falling, unbiased \pT jet spectra.

JETSCAPE also simulated jet quenching in a QGP. A hydrodynamically simulated QGP was generated for 3100 Au+Au collisions with 0-5\% centrality at \sNNcc. Each of these hydro events was hadronized 10 times and embedded with a unique jet each time, for a total of 31,000 jets quenched in hydro. Additionally, millions of quenched jet events were simulated using QGP bricks of various discrete thicknesses. A spectrum of $pp$ (i.e., non-quenched) jets was generated both as a baseline, and as a dataset used to train the neural networks. Finally, an additional spectrum of jets quenched in QGP bricks of length \SI{3.5}{fm} was generated.

\begin{figure}
    \centering
    \includegraphics[width=0.50\textwidth, trim=1.3cm 0.25cm 1.5cm  1.7cm, clip]{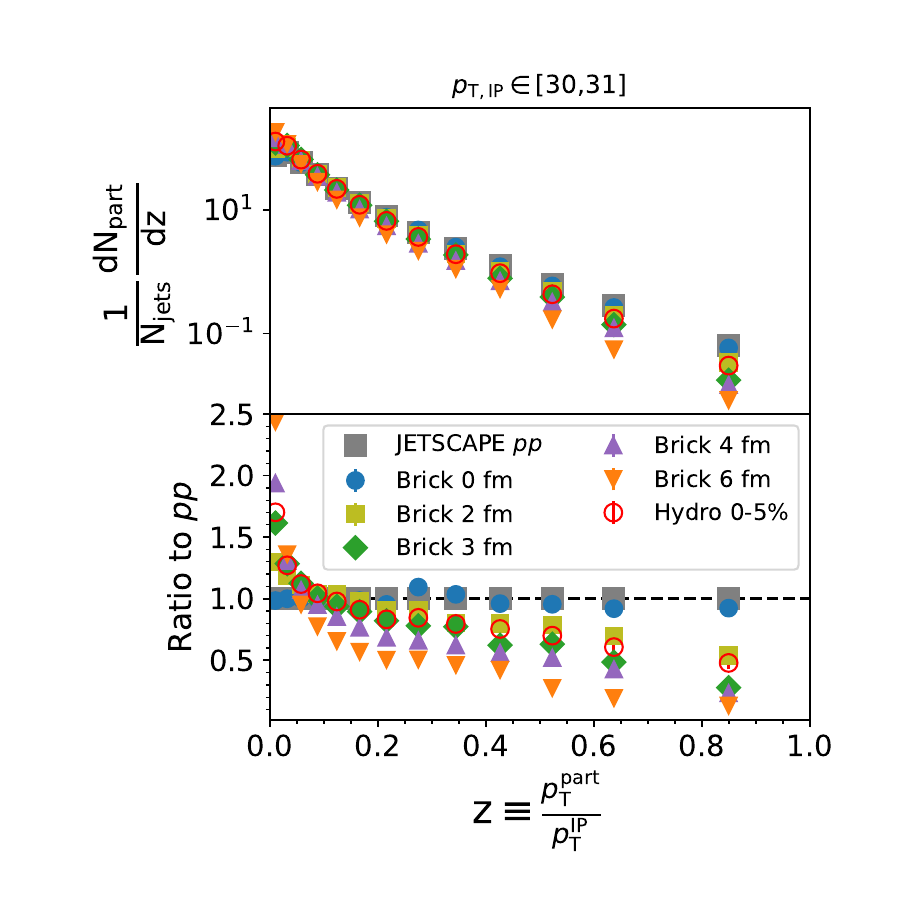}
    \caption{
    Distribution of the number of jet constituents ordered by their \pT ($p_\mathrm{T}^\mathrm{part}$) scaled by the \pT of the initiating parton ($p_\mathrm{T}^\mathrm{IP}$) for events with $\pThat\in[30,31] \;\mathrm{GeV}/c$. Statistical error-bars are mostly smaller than marker sizes.} \label{fig:FragFunc}
\end{figure}

The effects of jet quenching on the jet constituent particles' \pT is shown in Fig.~\ref{fig:FragFunc} for jets in $\pThat\in[30,31]\,\mathrm{GeV}/c$ events. The figure shows the jet fragmentation function, the average number of constituents per jet in increments of ``$\mathrm{Z}$'', the fractional constituent \pT relative to the \pT of the IP (as opposed to lower-case ``$z$'', in which the constituent \pT values are scaled by their fraction of the jet \pT instead of the IP \pT). The results show that the hydro events have quenching effects at low $\mathrm{Z}$ comparable to QGP brick lengths of 3-\SI{4}{fm}, and at higher $\mathrm{Z}$ comparable to 2-\SI{3}{fm}. 

\subsection{Background Production}

JETSCAPE identifies particles resulting from high-\pT scatterings from the background particles separately from those resulting from other processes. When clustered without the background particles, these result in ``truth jets''; when clustered with background particles, they result in ``reco-jets''. The hydro events generate realistic distributions of background particles, and therefore the set of background particles in each hydro event is saved into an external file. When processing the events with bricks of QGP (or without any QGP) only the particles of the leading truth jets are kept, while the background particles and other truth jets are discarded. The saved truth jet particles are embedded into a set of background particles from a hydro event before reconstructed as reco-jets.

The distribution  of the number of background of particles per hydro event is given in Fig.~\ref{fig:bulk_N}; distributions in $\phi$, $\eta$, and \pT are also given in in the Appendix in Fig.~\ref{fig:bulkphietapt}. 

\begin{figure}[h]
    \centering
    \includegraphics[width=0.50\textwidth, trim=0.3cm 0.30cm 0.0cm 0.0cm, clip]
    {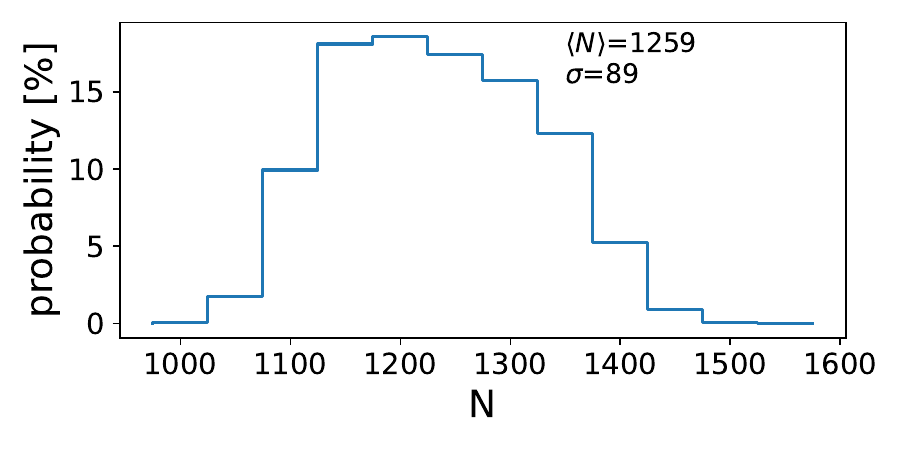}
    \caption{Distribution of the numbers of background particles per event.}
    \label{fig:bulk_N}
\end{figure}

\subsection{Jet Clustering and Matching}\label{sec:jetclustmatch}

The input data for each event consists of the set of particles associated with only the highest \pT initiating parton, which are clustered into the truth jets,  and the set of background particles generated from a hydro event. (For all jets generated in an hydro event, the jet is always clustered with the background particles from the same event). In each event, the following process was followed, utilizing FastJet \cite{Cacciari:2011ma} version 3.4.2 for all jet clustering with jet resolution parameter $R=0.4$.

\begin{enumerate}
    \item Use only the highest-\pT IP scattering for all hard scatterings in each event.
    \item Cut events with IP pseudorapidity $\eta^\mathrm{IP}>|1.0|$.
    \item Cluster all final-state particles resulting from the selected IP into \antikT jets. Consider all jets relative to the IP within distance $\left(\Delta{}R\equiv\sqrt{(\eta^\mathrm{jet}-\eta^\mathrm{IP})^2+(\phi^\mathrm{jet}-\phi^\mathrm{IP})^2}\right)$ of $\Delta{}R<0.4$. Discard the event if there are no such jets. If there are, select the highest-\pT of these jet as the truth jet with \pTtruth.
    \item Cluster the background particles into \kT jets \cite{Ellis:1993tq}. In hydro events, use the background particles from the same hydro event in the clustering. In all other events, use the background particles saved from one of the hydro events. 
    \item Remove the two highest \pT jets, and record the median jet-\pT density (\pTjet/\Ajet) as \rhobkg.
    \item Cluster the constituents from the IP scatterings and the background particles together  into \antikT jets. Select all resulting jets that are within $\Delta{}R<0.3$ of the truth jet. If there are none, discard the event. Otherwise, the highest-\pT of these jets is the ``reco jet'' with \pTreco.
    \item Record \pTtruth, \pTreco, and other event parameters used to train Neural Networks. A list of which parameters are used to train each neural network is given in Table~\ref{tab:NNinputs}.
\end{enumerate}

\begin{table}[h]
    \centering \caption{Neural network (NN) training parameters }\label{tab:NNinputs} 
    \begin{tabular}{p{0.18\linewidth}p{0.63\linewidth}}
    \toprule
     Label & Additional Training Parameters$^\dag$\\ \midrule
     \NN{AB} & (none) \\
     \NN{Ang} & Angularity: $\alpha\equiv\sum_i p_{\mathrm{T},i} \Delta{}R_{i}$,  where $i$ runs over all constituents, and $\Delta{}R_{i}$ is the $\eta$-$\phi$ distance from the constituent to the jet axis \\
     \NN{Ncons} & The number of jet constituents \\
     \NN{pTcons} & $p_\mathrm{T}$ of the highest-\pT constituents (limited to 10) in the reco jet \\
     \NN{AllReco} &  All parameters listed in this chart together \\
    \bottomrule
    \multicolumn{2}{c}{$^\dag$\pTtruth, \pTreco, \Ajet, \& \rhobkg, are used with each NN}\\\hspace{0.5em}
    \end{tabular}
\end{table}

\subsection{Neural Network Training}

The NNs were trained using the TensorFlow library \cite{Abadi:2016kic}. Each NN was composed of a sequential model using RELU activation functions with three dense layers of 100, 50 and 50 nodes respectively, with an additional final layer of a single node for the output value (\pTtruth). Each NN was trained with 12 epochs.

\begin{figure}[H]
    \centering
    \includegraphics[width=0.45\textwidth, trim=1cm 0.8cm 1cm 1cm, clip] {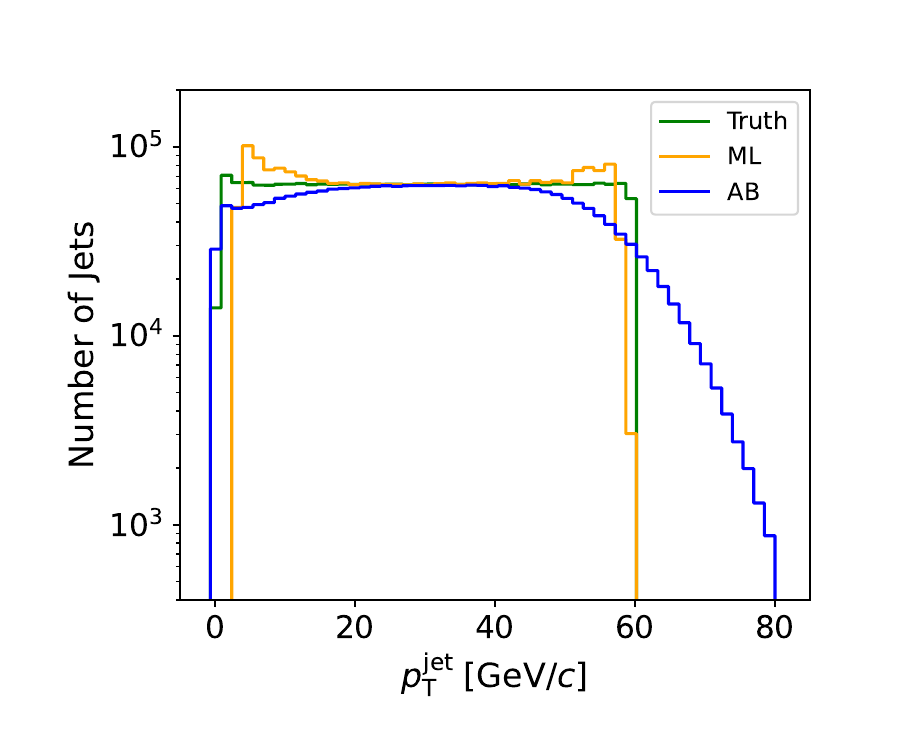}
    \caption{Jet \pT distributions: truth jet, reco-jet corrected by the \NN{AllReco} neural network, and the reco-jet corrected using the AB method ($\pTcorr\equiv{}p_\mathrm{T}^\mathrm{reco}-\rho\times{}A$).}
    \label{fig:truth_ml_rho_spectra}
\end{figure}

\begin{figure}[H]
    \centering
    \includegraphics[width=0.45\textwidth, trim=2.7cm 2cm 6.5cm 1.2cm, clip] {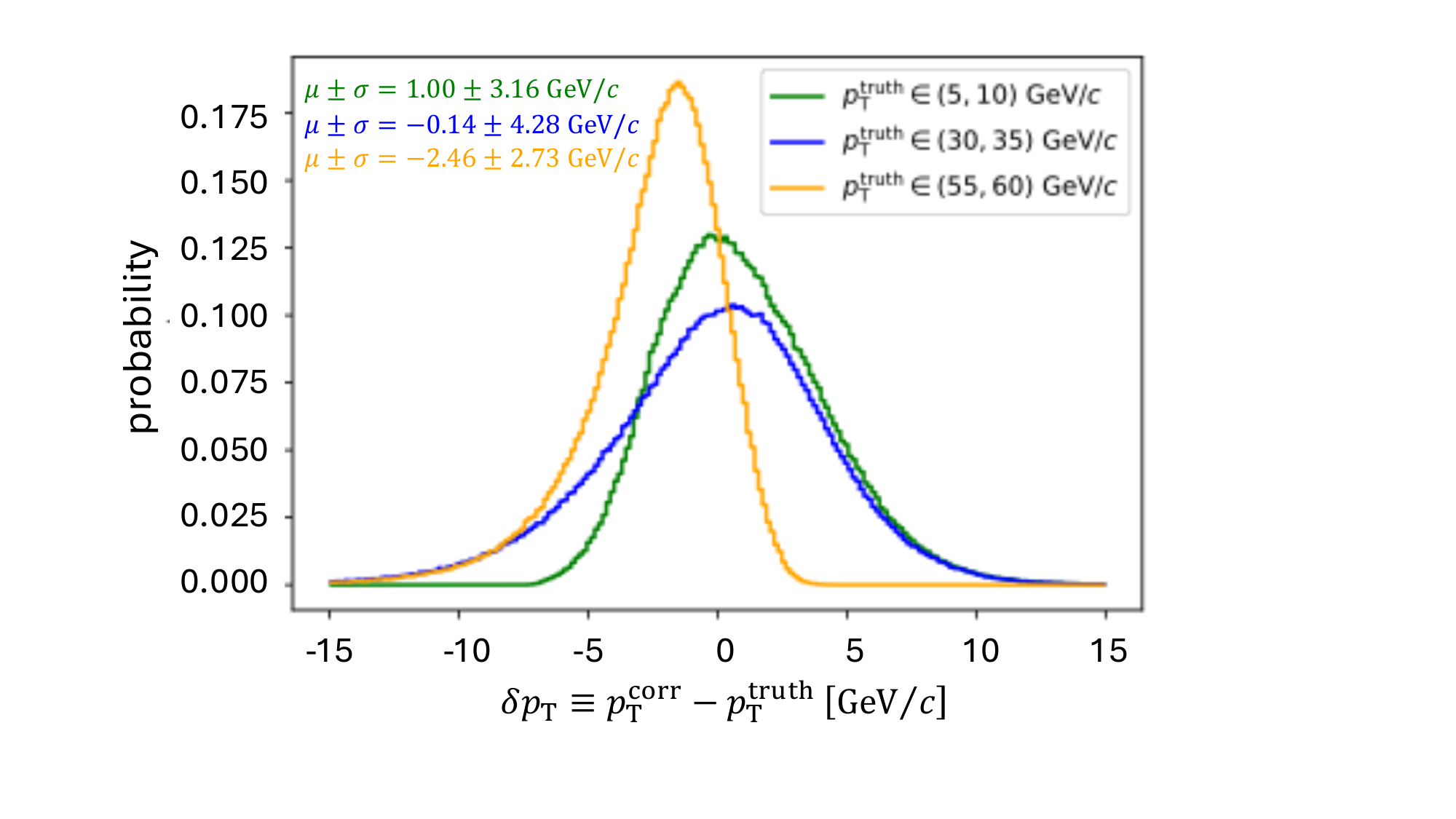}
    \caption{The probability distribution of the residual error, $\pTresid\equiv\pTcorr-\pTtruth$, with \pTcorr from \NN{AllReco} for events with three ranges of \pTtruth. Also listed are the mean and standard deviation of \pTresid for each range.}
    \label{fig:three_ptresid}
\end{figure}

In order to train the neural networks, an set of $pp$ jets with a flat \pT spectrum was generated, clustered, and embedded in backgrounds from hydro events, as described in Sec.~\ref{sec:jetclustmatch}. Each of the five NNs listed in Table~\ref{tab:NNinputs} were trained on this data set. The approximate importance of each parameter in each NN is visualized in Fig.~\ref{fig:gini_values}, which displays the feature importances using training random forest models trained with the same parameter lists.
Some results from \NN{AllReco} are given in Figures~\ref{fig:truth_ml_rho_spectra}-\ref{fig:three_ptresid}. Fig.~\ref{fig:truth_ml_rho_spectra} reports the \pTtruth spectrum, the \pTcorr spectrum generated by correcting \pTreco with \NN{AllReco} and also the \pTcorr spectra corrected using the AB method (without any NN).

Note that when applying the AB method any jet with $\pTcorr<\gevc{0}$ is discarded. This cut is commonly used in jet measurement analyses of real data. This results in the sharp boundary on the left hand side of the AB \pTcorr distribution in Fig.~\ref{fig:truth_ml_rho_spectra}. There is no cut on jets used to train the neural network; however, the neural network itself learns that there are no training jets with $\pTtruth\le\gevc{0}$ or $\pTtruth>\gevc{60}$. As such, the ML \pTcorr distribution has a sharp cut at both the left-hand and right-hand side side. This is to say, the ML has ``learned'' that any \pTreco value close to \gevc{0} cannot correspond to a truth jet sitting in an upward fluctuation in the bulk particle background; similarly any value close to \gevc{60} cannot correspond to a truth jet in an downward fluctuation. This learned constraint is an artifact of the training data and has no corresponding truth in the physics of actual experimental data: a jet with $\pTreco~\gevc{60}$ could be the result of a higher \pTtruth jet sitting on a downward background fluctuation. However, in data the importance of downward fluctuations is suppressed by the steeply falling \pTtruth spectra.

\begin{figure}[H]
    \centering
    \includegraphics[width=0.40\textwidth, trim=10.8cm 8.9cm 17.0cm 5.0cm, clip] {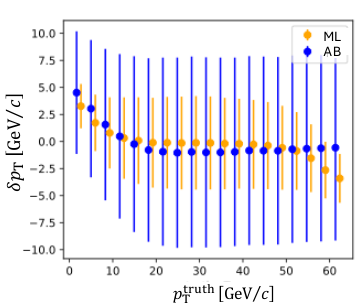}
    \caption{The values of the mean (markers) and standard deviation (length of the attached vertical lines) of the $\pTresid\equiv\pTcorr-\pTtruth$ distributions for events with bins of \pTtruth (as indicated on the $x$-axis), with \pTcorr calculated by the area-based method and by \NN{AllReco} (refer to Table~\ref{tab:NNinputs}). For visual clarity, the AB and ML markers are displayed with a small relative horizontal offset.}
    \label{fig:meanstd_cont}
\end{figure}

Because the NNs learn the boundaries, the \pTresid values near those boundaries are highly biased. To illustrate this effect, Fig.~\ref{fig:three_ptresid} shows the \pTresid distributions for events with three ranges of \pTtruth: one near each boundary and one in the middle. Note the ``inward-tales'' of the distributions near the boundary agree with the tails of the middle distribution, while their ``outward'' tales are truncated. The figure also reports the mean and standard deviation of each \pTresid distribution. In Fig.~\ref{fig:meanstd_cont} the mean and standard deviations of \pTresid for a  continuous set of ranges of \pTtruth are shown; markers show the mean \pTresid values, and the error bars are scaled to equal the magnitude of the \pTresid standard deviations. As is visually apparent, the neural network would introduce significant off-diagonal entries near the training boundaries in a correlation matrix of \pTtruth and \pTcorr, whereas the AB method, while having overall significantly larger distributions, would only do so for low-\pT jets.

\section{Neural Network Performance}

\subsection{Effects of Jet Quenching on \pTresid from Neural Networks}\label{subsec:jet_quench_errors}

\begin{figure}[H]
    \centering
    \includegraphics[width=0.45\textwidth, trim=0.3cm 0.3cm 0.3cm 0.3cm, clip] {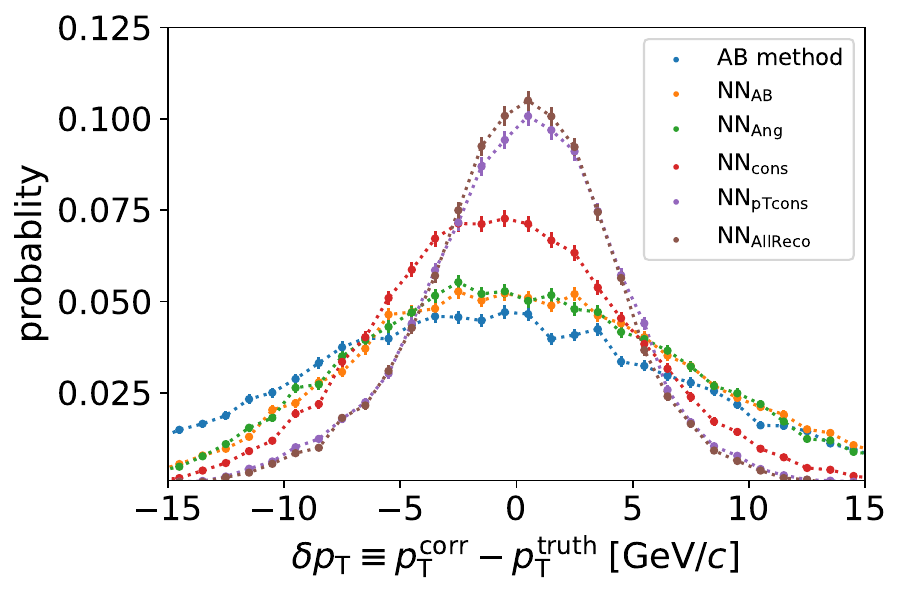}
    \caption{The residual error distribution in neural network corrections from \pTreco to \pTcorr for events generated with $\pThat\in[30,31]\;\mathrm{GeV/\mathit{c}}$ and trained on parameters listed in Table~\ref{tab:NNinputs}.}
    \label{fig:NNcomped}
\end{figure}

Figure~\ref{fig:NNcomped} shows the \pTresid distributions from use of the five NNs listed in Table~\ref{tab:NNinputs} in events with $\pThat\in[30,31]$ of $pp$ jets. The same results for four other ranges of \pThat are given in the appendix in Fig.~\ref{fig:NNcomp_allpthat}.

\begin{figure}[H]
    \centering
    \includegraphics[width=0.45\textwidth, trim=0.8cm 0.80cm 1.5cm 1.1cm, clip] {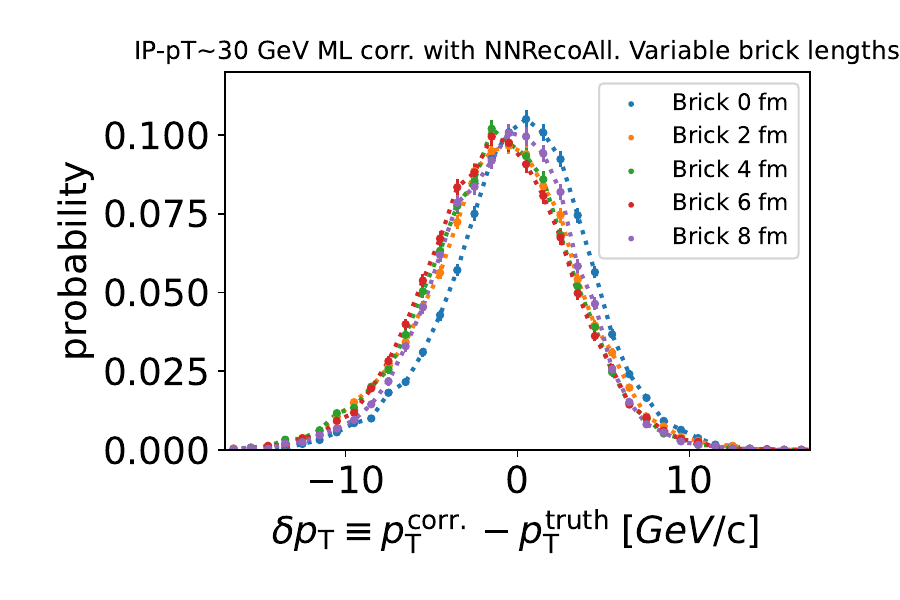}
    \caption{The evolution of the distribution of $\pTresid\equiv\pTcorr-\pTtruth$ for jets undergoing quenching in different lengths of QGP (as listed) in which \pTcorr generated by \NN{AllReco} (refer to Table~\ref{tab:NNinputs}) which was trained on jet events with no quenching. The events were generated with $\pThat\in[30,31]\;\mathrm{GeV/\mathit{c}}$.}
    \label{fig:one_panel_fm_evolve}
\end{figure}

When the NN background correction is performed on quenched jets, the \pTresid distributions evolve with the magnitude of quenching. This evolution is shown in Fig.~\ref{fig:one_panel_fm_evolve} for \pTresid for jets in QGP bricks from \SI{0}{fm} (no quenching) up to \SI{8}{fm} using \NN{AllReco}. Comparable results for all the NNs, for five different selections of \pThat, are given in Fig.s~\ref{fig:resid_fm_prog_rhoA}-\ref{fig:resid_fm_prog_recoall} in the appendix. Note that the biggest shift in \pTresid occurs within the first two fm of quenching, which is within the range of quenching experienced by the jets in the hydro events and expected at RHIC energies.

\begin{figure}[H]
    \centering
    \includegraphics[width=0.45\textwidth, trim=0.4cm 0.80cm 0.6cm 2.0cm, clip] {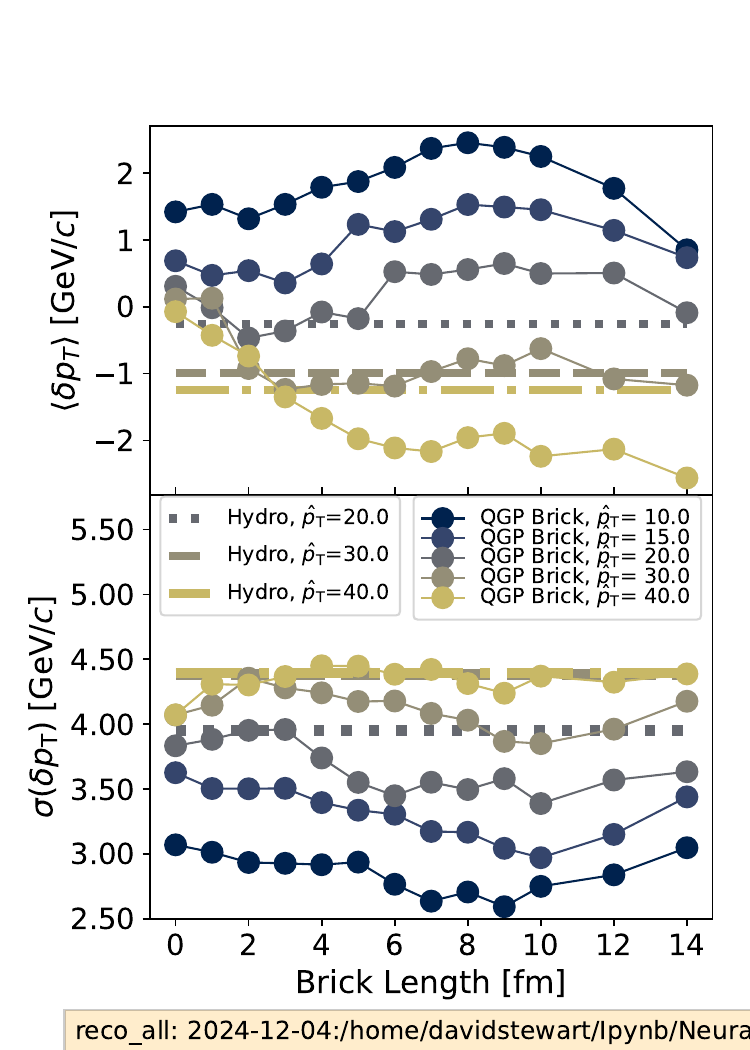}
    \caption{The mean and standard deviation of the $\pTresid\equiv\pTcorr-\pTtruth$ distributions for \pTcorr generated from neural network \NN{AllReco} (trained with all parameters listed in Table~\ref{tab:NNinputs} row (e)). Jets where quenched using a hydrodynamically simulated QGP, as well as bricks of QGP, and  no quenching (at brick length of 0). The hydro data aren't associated with set QGP brick lengths, and are therefore displayed with horizontal lines located vertically at their $\langle\pTresid\rangle$ and $\sigma(\pTresid)$ values.}
    \label{fig:mean_std_prog_reco_all}
\end{figure}

In order to compare the quenching effects between hydro and brick events, the mean ($\langle\pTresid\rangle$) and standard deviation ($\sigma\left(\pTresid\right)$) of each \pTresid distribution is reported for a series of brick lengths, along the values for hydro events. In the hydro events, the jet quenching path lengths are modeled individually in each event, and as such are not associated with a specific path length. Therefore, the hydro $\langle\pTresid\rangle$  and $\sigma(\pTresid)$ values are displayed as horizontal lines. These are shown in Fig.\ref{fig:mean_std_prog_reco_all} for jets at $\pThat\in[30,31]\;\mathrm{GeV}/c$ using \NN{AllReco}. The corresponding figures for the other four neural networks are given in the appendix in Fig.~\ref{fig:mean_std_prog_rhoA_and_angle} and Fig.~\ref{fig:mean_std_prog_nconst_and_cpt}. In each case, the biases in \pTresid for hydro events correspond to those in events using QGP bricks of 3 to \SI{4}{fm}, which is consistent with the modification in jet fragmentation shown in Fig.~\ref{fig:FragFunc}, and confirms the choice of using quenching in \SI{3.5}{fm} bricks of QGP as proxies for quenching in hydro events.


\subsection{Effects of \pTresid Bias on Measuring \pTjet}

As seen in Fig.~\ref{fig:mean_std_prog_reco_all}, using NNs for background corrections biases the resulting distributions of \pTresid. Of particular note, the biases are also jet-\pT dependent. In order to demonstrate the magnitude of effects these biases can result in actual measurements, we simulate a measurement of jet quenching using the NNs.  The algorithm employed parallels the process used in actual detector jet measurements, and is simplified to focus on the jet background rejection biases introduced by the mismatch in jet substructure between the quenched jets and those used to train the NNs. An outline of the methodology used in an actual experiment is given in Sec.~\ref{sec:alg_exp}, while the differences in this paper are listed in Sec.~\ref{sec:alg_sim}.

\subsubsection{Algorithm to Measure Jets Quenching in Experiments}\label{sec:alg_exp}
\begin{enumerate}[label=\textbf{\alph*.}] 
\item\label{it:0}
Measured data consists of events with jets constituents clustered together with the heavy background resulting from heavy ion collisions. This clustering results in detector-level jets, with a spectrum of \pTreco (i.e. $\mathrm{d}\pTreco/\mathrm{d}\pT$).
\item\label{it:1} The reco-jets are background corrected, commonly using the AB method, to jets with \pTcorr.
\item\label{it:2} To determine the ``truth-level'' jets -- i.e. the jets which would result if only the particle-level jet constituents were clusters -- generate a correlation matrix correlating \pTtruth to distributions of \pTcorr. To do this:
\begin{enumerate}[label=\textbf{c.\arabic*}]
\item\label{it:3}
Use a MC generator to simulate $pp$ jets (with corresponding values of \pTtruth).
\item Propagate the $pp$ jet constituents through a detailed physics simulation of the detector (e.g., \cite{GEANT4:2002zbu}) to generate detector-level final state hadrons.
\item\label{it:4}
Separately, collect distributions of background hadrons from actual collision measurements. These are really just the measurements of evens with minimal trigger requirements.
\item\label{it:5} Embed the detector-level hadrons of the $pp$ jets into the actually measured backgrounds, and cluster into ``reco-jets''.
\item\label{it:6} Background correct the reco-jets into corr-jets using the same method as used in the measurement in step \ref{it:1}
\item\label{it:7} Geometrically match the simulated $pp$ jets (``truth jets'') to the corr-jets, using some cutoff in 
$\Delta{}R$, 
$\left(\Delta{}R\equiv\sqrt{(\eta^\mathrm{truth}-\eta^\mathrm{corr})^2+(\phi^\mathrm{truth}-\phi^\mathrm{corr})^2}\right)$. Fill pairs of matched jets into response matrix \Mct.
\begin{itemize}[noitemsep,topsep=0pt]
\item Unmatched $pp$ jets are counted as ``misses'' and account for inefficiency in jet detection.
\item Unmatched reco-jets are ``fakes'' and account for jets resulting from clustering only the background.
\end{itemize}
\end{enumerate}
\item\label{it:8}
Use \Mct along with the \pT spectra of the misses and fakes, to statistically correct the detector measured \pTcorr to the measured \pTtruth.
\item\label{it:9} Scale the A+A events to an equivalent number of $pp$ collisions. This is because hard partonic scatterings (jets) scale in a heavy ion collision by the number of equivalent nucleon-nucleon collisions; e.g., for hard scatterings, a head-on Au+Au collision is the equivalent of upwards of 1000 $pp$ collisions. Report the ``nuclear modification factor'' \RAA, which is ratio of the scaled jet spectra to that in $pp$; so named because if there is no jet quenching, then $\RAA\approx1.$
\end{enumerate}

\subsubsection{Algorithm Used in this Paper to Measure Jets}\label{sec:alg_sim}

The algorithm implemented for the results reported in this paper are comparable to that listed in Sec.~\ref{sec:alg_exp}. The differences listed below.

\begin{enumerate}[label=\textit{\alph*.}]
\item\label{ii:0} The ``measurement data'' consists of JETSCAPE simulated jets quenched in \SI{3.5}{fm} bricks of QGP which are embedded into background collected from the JETSCAPE hydro events. The following cuts are applied:
\begin{itemize}
    \item For the jets, use only jets from the highest-\pT IP collision.
    \item If that IP has $|\eta|>1$ discard the event.
    \item Cluster the hadrons from this IP into \antikT the jet(s) with $R=0.4$ jet using \texttt{FastJet} 3.4.2 \cite{Cacciari:2011ma}.
    \item Find the highest-\pT reconstructed jet within $\Delta{}R<0.4$ of the IP, and label it as the ``truth-jet''.
    \item Embed the constituents of the truth-jet into the background hadrons from a hydro event, and re-cluster into ``reco-jets''.
    \item Find all reco-jets within $\Delta{}R\le0.3$ of the truth-jet, and save the highest-\pT jet as the reco-jet.
\end{itemize}
Herein, this process maps only the single ``leading jet'' from the highest-\pT IP. This is done for simplicity in the modeling.  In an actual measurement (e.g., \ref{it:0}), of course, there is no knowledge what the underlying jet distribution is, and all jets per event would be recorded. As lower-\pT jets experience the effects of quenching more strongly, this method likely underestimates the final results of the effects of quenching.
\item Same as \ref{it:1}, in which the background correction is conducted six times, once with the AB method and once with each of the five NNs (as listed in Tab.~\ref{tab:NNinputs}). The results of all six methods are separately propagated through the following steps to individual results.
\item The steps to generate the \Mct are identical to those in \ref{it:3}-\ref{it:7}, with the following specifics:
\begin{enumerate}[label=\textit{c.\arabic*}]
\item The embedded jets are the leading JETSCAPE $pp$ jets, using the same cuts as in Step~\ref{ii:0}
\item Besides fiducial cuts in rapidity,  $|\eta|\le1.0$ for IP and $|\eta|\le1.1$ for hadrons, the simulation does not account for detector effects. This is done to focus only on the effects of the NNs.
\item The background consist of the same body of hydro events as used in Step~\ref{ii:0}
\end{enumerate}
\item[] \textit{c.4-c.6} Identical to Steps~\ref{it:5}-\ref{it:7}.
\item The correction using \Mct is done using an iterative Bayesian unfolding procedure implemented with the RooUnfold package \cite{Adye:2011gm}, as is also used in actual measurements. Alternatively, the correction is also done by using single bin efficiencies (``1-Bin Eff.'' in the result labels), in which the ratio of $\pTtruth/\pTcorr$ in the simulation is used as a single scaler multiplier to correct each \pT bin of the corrected jet spectra.
\item As this procedure is already scaled to only the leading jet generated per nucleon-nucleon collision (the quenched jets are generated as $pp$ jets which are quenched in \SI{3.5}{fm} bricks of plasma which are then embedded into backgrounds), then ``\RAALeadJet'' is defined directly as the ratio the measured jet spectra to a truth spectrum of leading truth jets in unquenched $pp$ events.
\end{enumerate}

\section{\RAALeadJet Results}

\begin{figure}[h]
    \centering
    \includegraphics[width=0.47\textwidth, trim=1.0cm 1.7cm 1.3cm 0.3cm, clip] {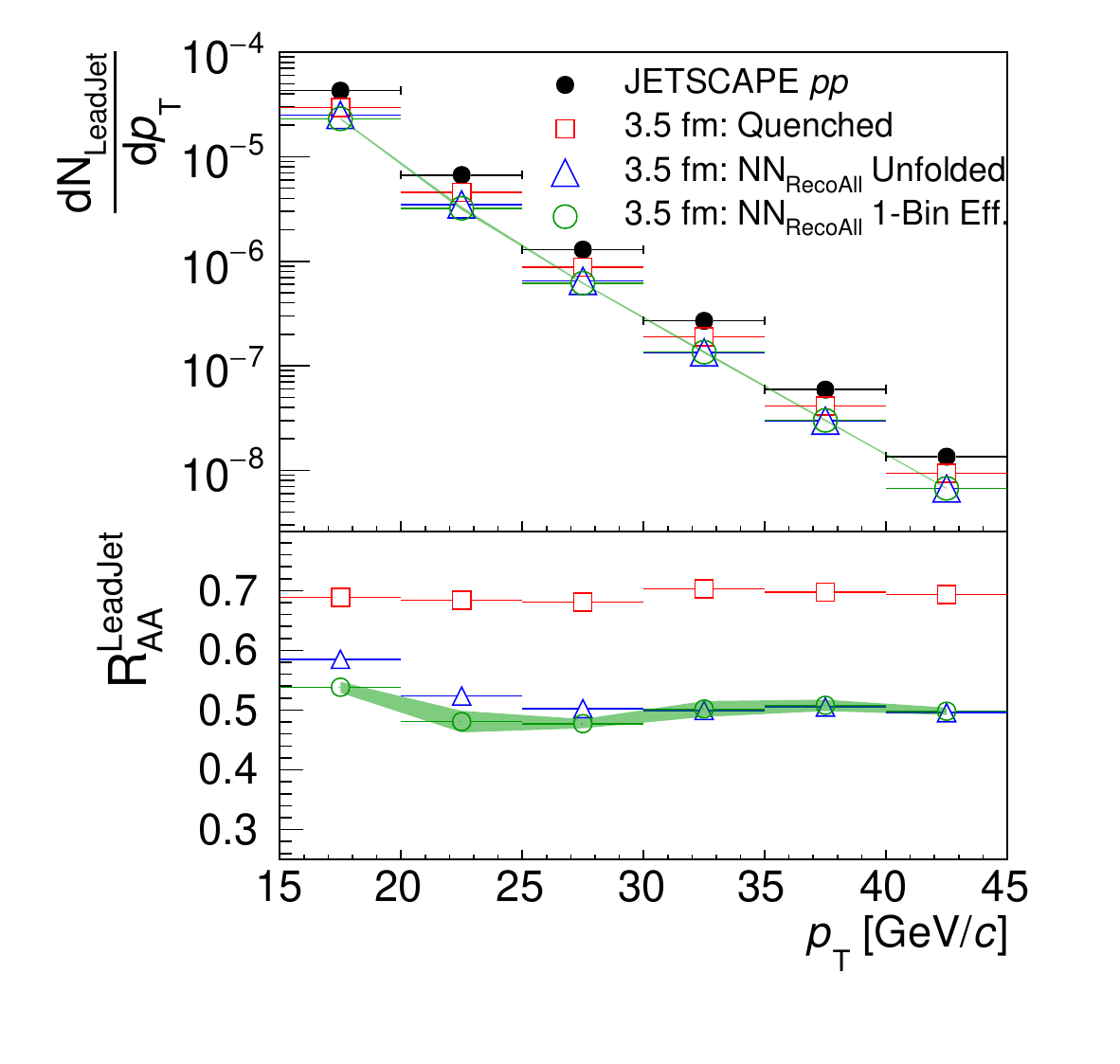} \caption{Top panel: Spectra of the leading jet per event for $pp$ (non-quenched) and quenched jets, counting only events with $\pTtruth>\SI{12}{GeV/\mathrm{c}}$ and $\pTreco-\Ajet\rhobkg>\SI{0}{GeV/\mathit{c}}$, in addition to the measured spectra of quenched jets which are background corrected using the $\mathrm{NN}_\mathrm{reco\;all}$ and unfolded using both the single-bin efficiency (``1-Bin Eff.'') or Bayesian unfolding with 4 iterations (``Unfolded''). Bottom panel, the $R_\mathrm{AA}$ of the actual quenched spectra and the measured quenched spectra. The shaded area (which is quite narrow in the top panel) represents the uncertainty introduced by instability in the unfolding. Statistical errors are smaller than the plotting markers.}
    \label{fig:RAA_RecoAll}
\end{figure}

The algorithm detailed in Sec.~\ref{sec:alg_sim} was used to calculate the corrected jet spectra for quenched jets using the AB method and all five NNs. The jet spectra for the unquenched jets ($pp$), truth-level quenched jets, and the ``measurement'' of the quench jet spectra are reported in Fig.~\ref{fig:RAA_RecoAll}, in which \NN{AllReco} was used for the jet background correction. The \RAALeadJet values for both the real and the measured \pT's are also shown. The same information is shown for the other four NNs in the appendix in Fig.~\ref{fig:RAA_like_Ang_Ncons} and Fig.~\ref{fig:RAA_like_cpt_AB}. For these jets -- the lead jet only per event quenched in \SI{3.5}{fm} bricks of QGP -- the actual \RAALeadJet values, as estimated using bricks of QGP in simulation, are approximately independent at around 0.7 on jet \pT, while the error introduced by using the NN on quenched jets resulted in \RAALeadJet values up to 30\% lower.

\begin{figure}[h]
    \centering
    \includegraphics[width=0.50\textwidth, trim=0.1cm 0.25cm 1.4cm 0.7cm, clip] {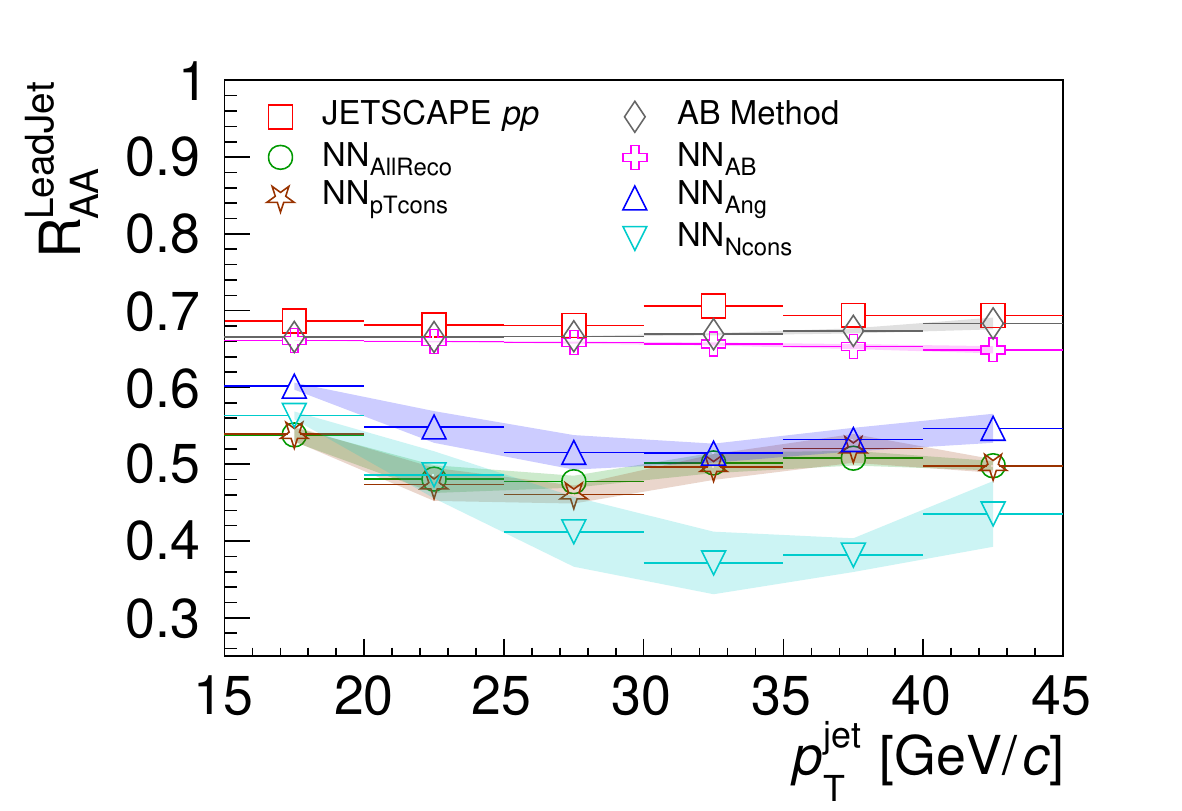}
    \caption{The measured $R_\mathrm{AA}$ of the quenched jets, background corrected with the traditional AB Method ($\pTreco-\Ajet\rhobkg$) and five neural networks, trained on parameters listed in Table~\ref{tab:NNinputs}. The actual ratio of the quenched jets to unquenched jets is also shown. Note that in this context, the spectra are only of the leading jet in each event, counting only events with the jet with $\pTtruth>\SI{12}{GeV/\mathit{c}}$ and $\pTreco-\Ajet\rhobkg>\SI{12}{GeV/\mathit{c}}$. The shaded areas account for the instability in the unfolding. Statistical errors are smaller than the markers.}
    \label{fig:RAA_summary}
\end{figure}

For comparison, the \RAALeadJet values resulting from using the AB method and all five NNs are shown together in Fig.~\ref{fig:RAA_summary}. Notably, the AB method results in approximately the true value, as does \NN{AB} which is trained on the same parameters use in the AB method. All the other NNs result in systematically lower \RAALeadJet values. Perhaps not surprisingly, \RAALeadJet generated with \NN{Ang} is the least biased out of these four NNs. If a jet undergoes partonic energy loss, and the jet clustering recovers (partly) the medium induced gluon emissions, then the impact on the angularity training parameter $\sum_i p_{\mathrm{T},i} \Delta{}R_{i}$ is mitigated. On the other hand, quenching increases the number of constituents monotonically and the corresponding bias in \RAALeadJet from \NN{Ncons} is the largest.


\section{Summary and Conclusions}

Models can be trained using substructure of jets clustered together with background particles from heavy ion collisions -- in addition to the median background \pT density -- to correct for the background component of the jet \pT, and result in considerably smaller residual errors than just subtracting the mean background in each event. However, in events with quenching, jet substructures are modified, which in turn biases the background corrections. This report studies those biases using best-in-class simulations from JETSCAPE to generate hydrodynamically modeled QGP, with resulting background particle distributions, and associated jet quenching in Au+Au collisions at \sNNcc. It should be noted, however, that jet-medium interactions are not captured in the simulations used.

We trained NNs on using a variety of jet substructure parameters using $pp$ jets embedded into the realistic heavy ion backgrounds, and reported the biases through the residual errors on background correction on quenched jets. The biases are observed to be significant. They are also dependent on jet-\pT, which is qualitatively different from those from AB background corrections which are independent of jet substructure whose resolution is limited by the inhomogeneity in the background density.

In order to demonstrate the possible magnitude of the error introduced by these biases, we mimic as closely as possible an actual jet \RAA measurement using the NNs for background \pT corrections. In order make these measurements, we compared quenching in computationally cheaper ``bricks'' of QGP, and found that bricks \SI{3.5}{fm} thick are a good proxy for quenching in collisions with hydrodynamically modeled QGP at RHIC energies. Accordingly, we simulated the \RAA measurement using jets quenched in these bricks of QGP and embedded into the backgrounds from the ``hydro'' events. The resulting error propagated in the \RAA are significant, up to a maximum of around 47\% when using \NN{Ncons}, and no less than 18\% for any \pT range for any NN. The only exception is \NN{AB} which is trained only on the parameters used in the AB method, and therefore independent of jet substructure.

Any application of background correction using jet fragmentation must presuppose an amount of jet quenching in order to then proceed to actually measure the jet quenching. It may be possible to parameterize the biases and determine bounding errors as has been done at LHC energies; see, e.g. \cite{ALICE:2023waz}. If, when such a procedure is used, it is not apparent which among the outcomes are the most probable, we strongly recommend the results be reported as a bounded range without any markers. Alternatively, it may also yet prove possible to use an iterative method to appropriately select and refine the value of the modeled quenching with the measurement of jets themselves.

Alternative to training NNs directly on the jet substructure, with the inherent uncertainties introduced by jet quenching, it may be possible to use ML to distinguish between jet-like objects which are clustered purely from background particles (i.e. ``fake jets'') and those containing real jet constituents. There is an essentially unbounded supply of background measurements available at each detector, and perhaps ML could become very sensitive to the presence of ``some signal'' in this background without being selectively sensitive to the type (or substructure) of that signal. If achievable, such a classifier could dramatically decrease the abundance of fake jets to real jets in measurements. In turn, this could facilitate jet measurements down to lower-\pT ranges than currently accessible as RHIC energies.

\section*{Acknowledgments}

This work was supported by the U.S. Department of Energy Office of Science, Office of Nuclear Physics under Award No. DE-FG02-92ER40713.
The authors would like to thank Hannah Bossi, Helen Caines, and Raymond James for insightful discussions on machine learning in jet physics, and Chun Shen for discussion and expertise in JETSCAPE.


\clearpage
\onecolumngrid
\appendix
\pagenumbering{roman}
\renewcommand{\thefigure}{A.\arabic{figure}}
\setcounter{figure}{0}

\section{Additional Figures}

\begin{figure}[h]
    \centering
    \includegraphics[width=0.60\textwidth, trim=0.3cm 1.80cm 1.0cm 2.0cm, clip]
    {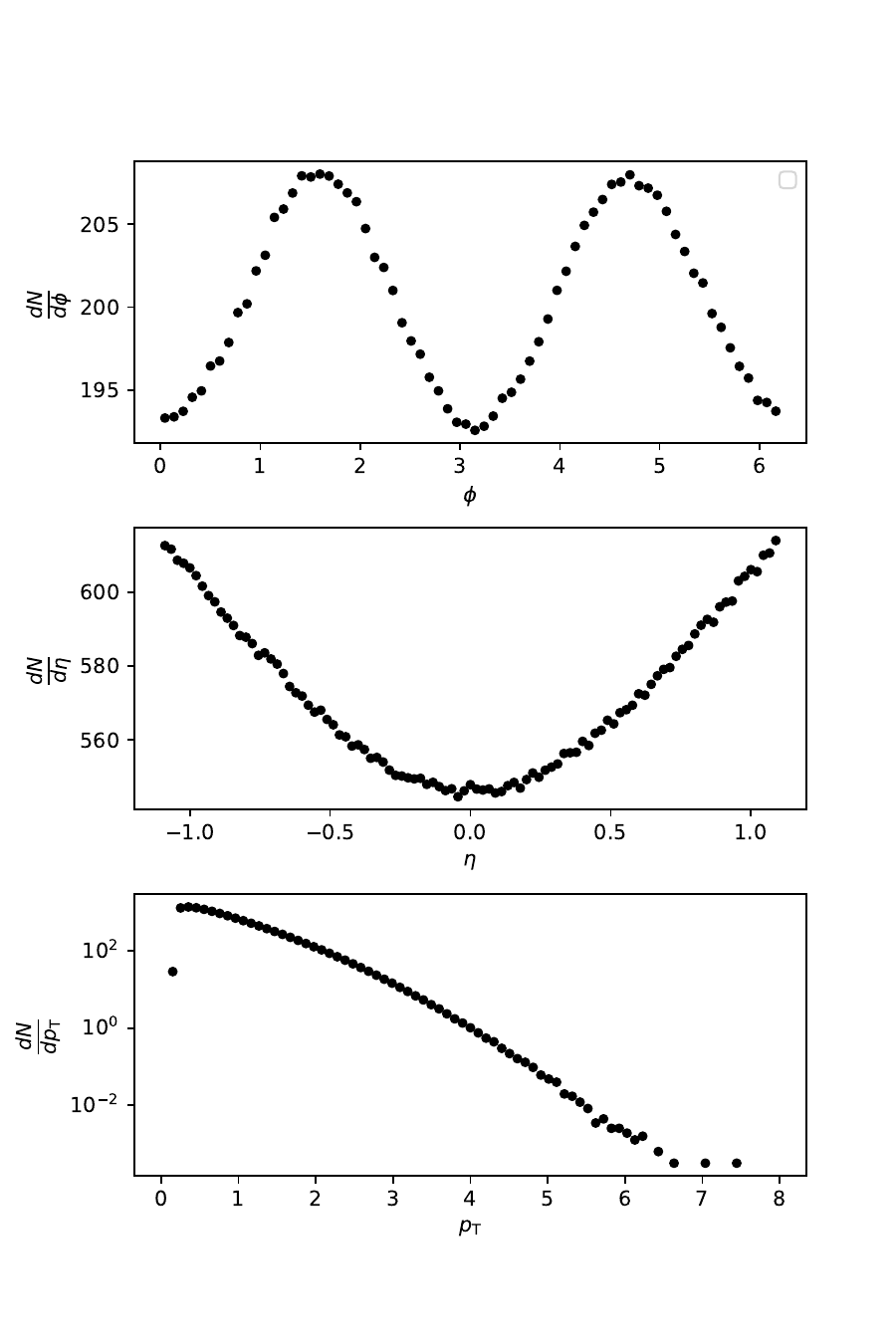}
    \caption{Distributions of $\phi$, $\eta$, and \pT densities of background particles, averaged over all background events. The $v^{2}$ flow shown is a result of JETSCAPE aligning all impact parameters along the $x$-axis.}
    \label{fig:bulkphietapt}
\end{figure}

\begin{figure}
    \centering
    \includegraphics[width=0.85\textwidth, trim=.0cm .0cm 0.5cm 1.0cm, clip]
    {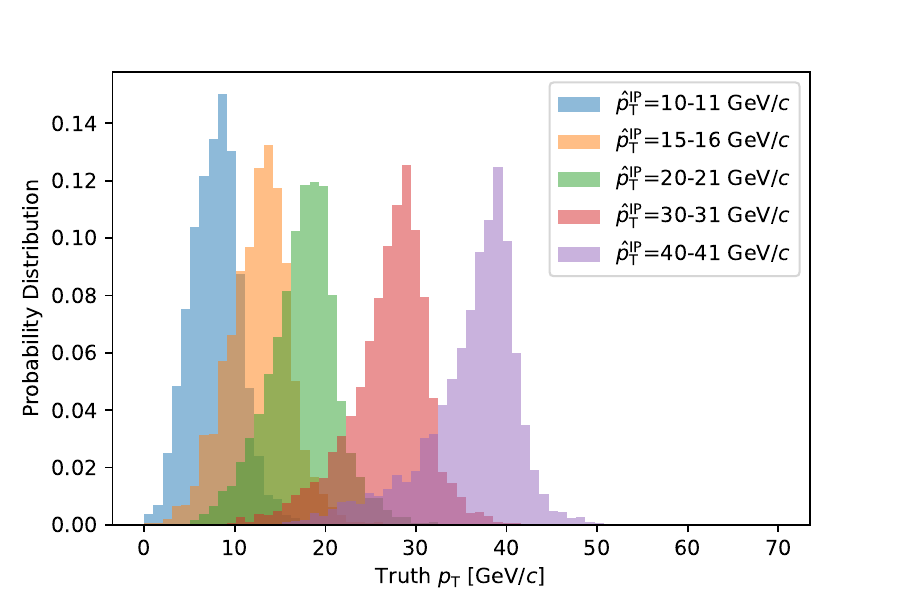}
    \caption{Distribution of \pTtruth per JETSCAPE \pThat parameter selection for the non-quenched datasets.}
    \label{fig:pTtruthPerIP}
\end{figure}

\begin{figure}
    \centering
    \includegraphics[width=0.75\textwidth, trim=0.5cm 0.80cm 1.9cm 0.7cm, clip] {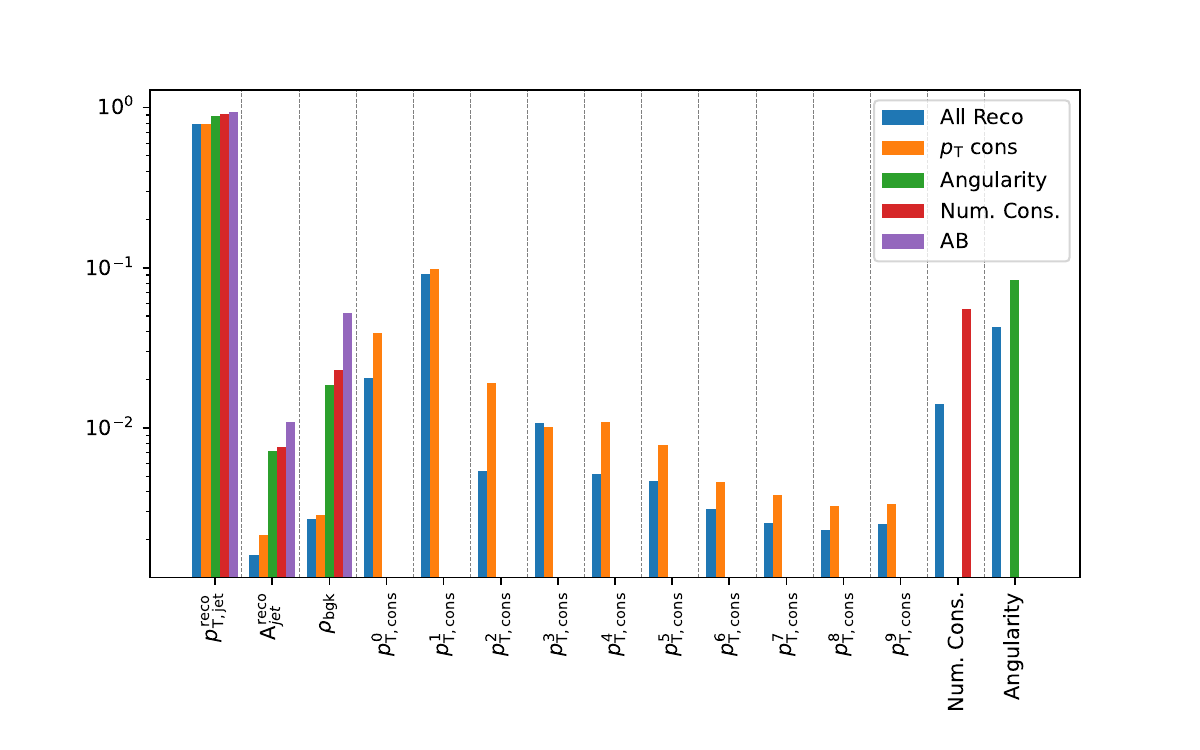}
    \caption{Random forest feature importances, using random forests trained on the sets of parameters (listed in Table~\ref{tab:NNinputs}) which were used to train the Neural Networks. These are consistent with those reported in prior publication \cite{Haake:2018hqn}.}
    \label{fig:gini_values}
\end{figure}

\begin{figure}[H]
    \centering
    \includegraphics[width=0.8\textwidth, trim=0.1cm 0.20cm 0.0cm 0.0cm, clip] {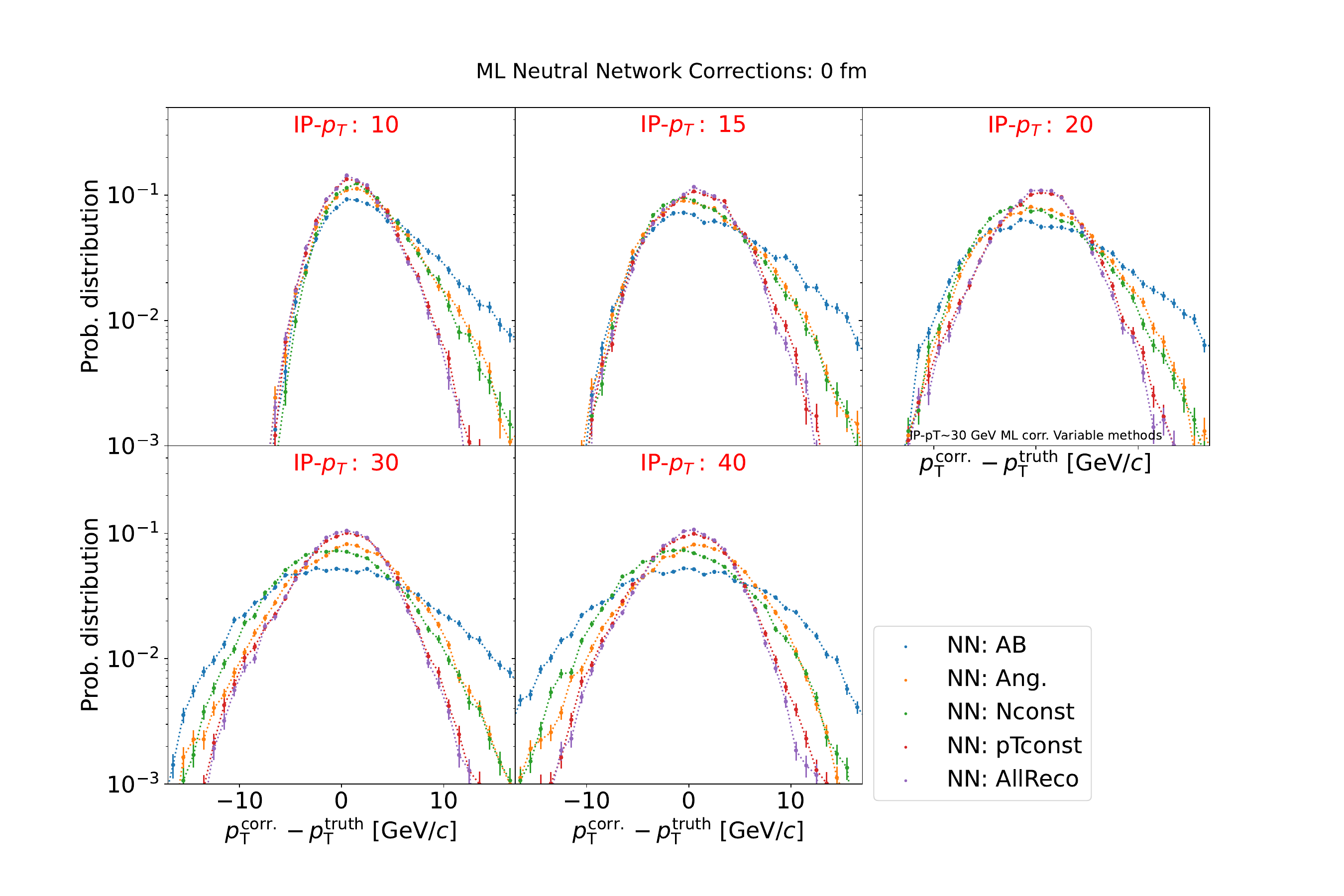}
    \caption{The distributions of $\pTresid\equiv\pTcorr-\pTtruth$ for distributions in neural network corrections from \pTreco to \pTcorr for events generated with five ranges of $\pThat$ and trained on parameters listed in Table~\ref{tab:NNinputs}.}
    \label{fig:NNcomp_allpthat}
\end{figure}

\begin{figure}[H]
    \centering
    \includegraphics[width=0.8\textwidth, trim=0.1cm 0.20cm 0.0cm 0.0cm, clip] {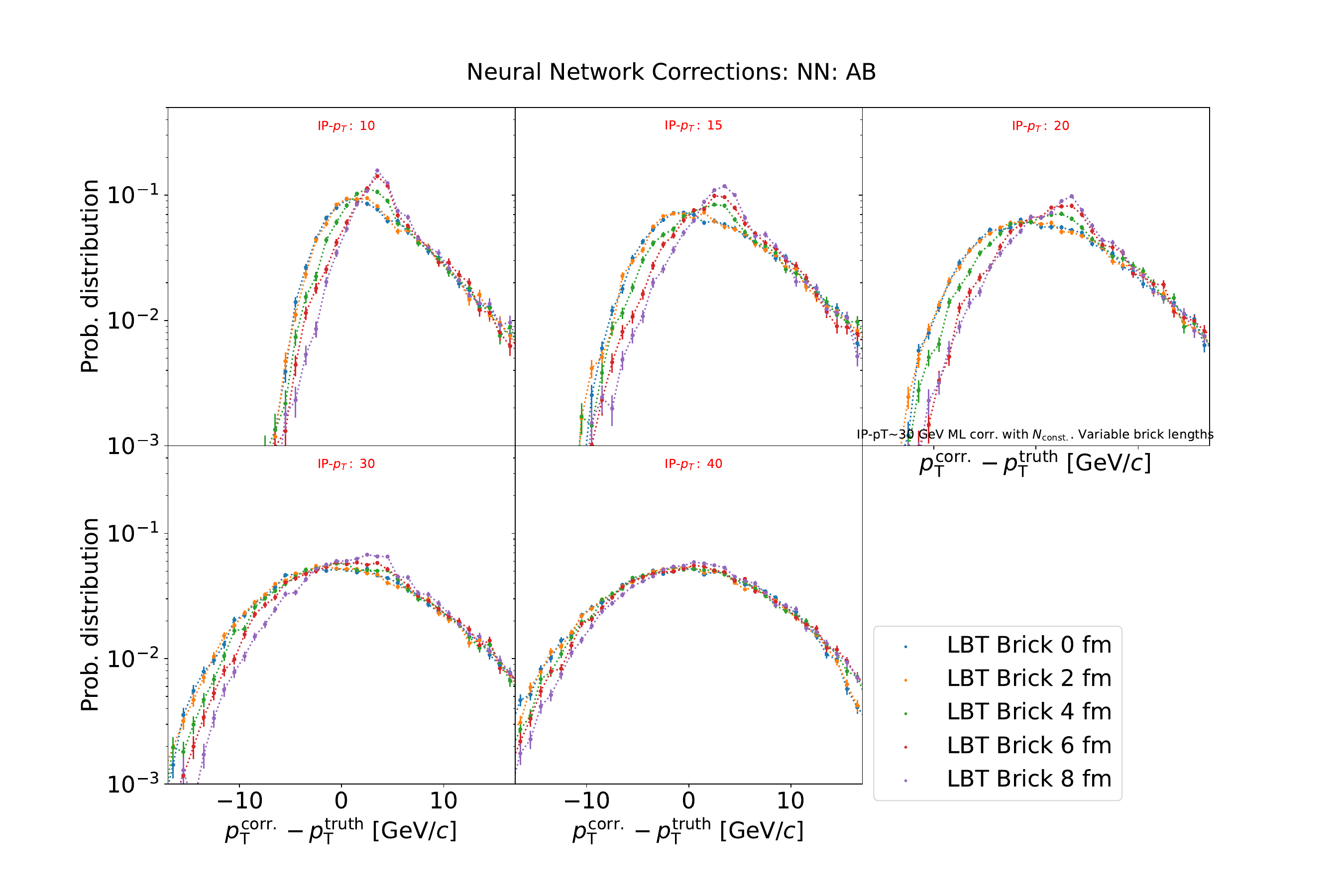}
    \caption{Distributions of $\pTresid\equiv\pTcorr-\pTtruth$ for distributions with correction made by a neural network trained on \rhobkg, \pTtruth, \pTtruth, and $A_\mathrm{jet}$. Within each panel, the evolution of \pTresid for events with no quenching to quenching with a \SI{8}{fm} brick of QGP.}
    \label{fig:resid_fm_prog_rhoA}
\end{figure}

\begin{figure}[H]
    \centering
    \includegraphics[width=0.8\textwidth, trim=0.1cm 0.20cm 0.0cm 0.0cm, clip] {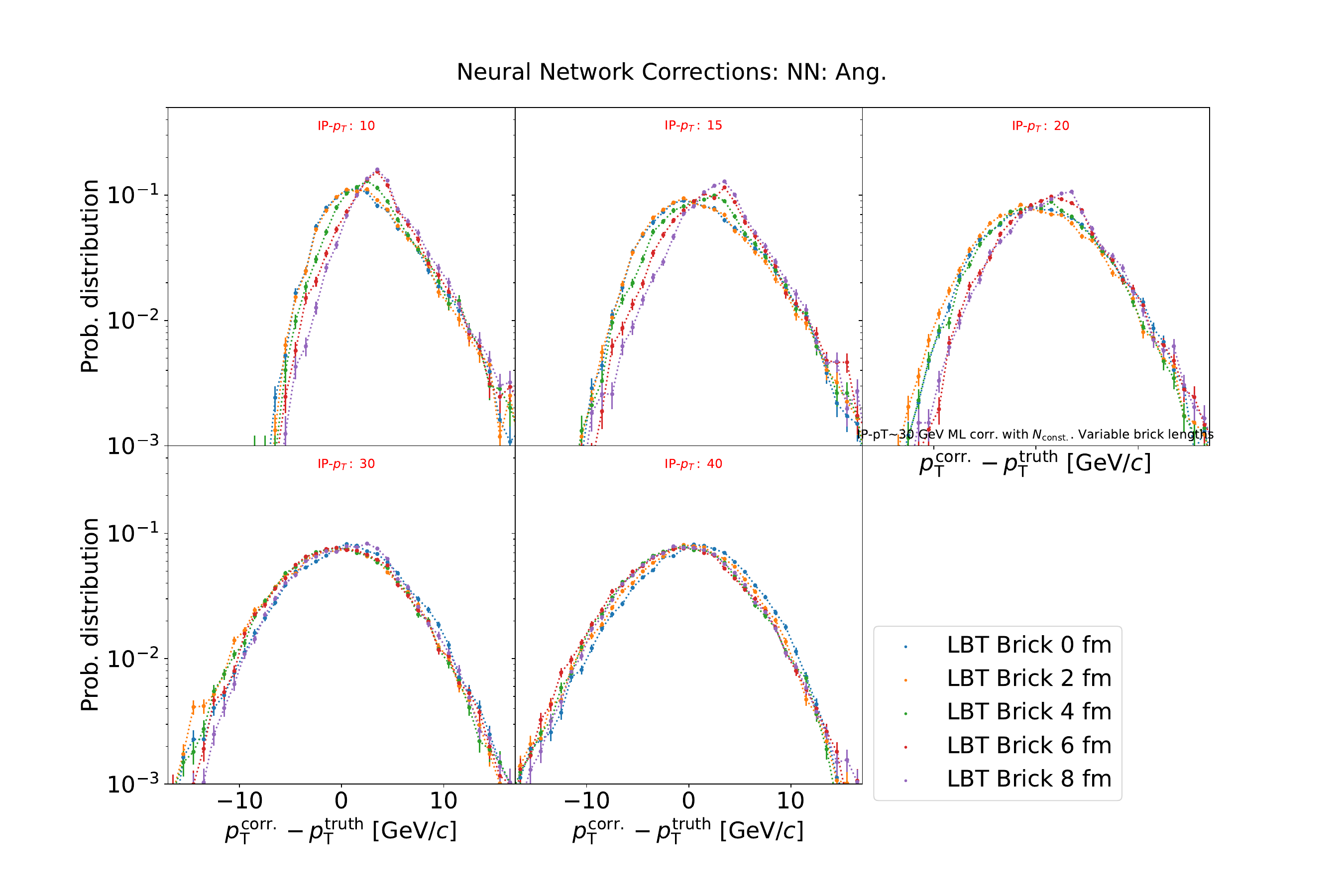}
    \caption{Distributions of $\pTresid\equiv\pTcorr-\pTtruth$ for distributions with correction made by a neural network trained on \rhobkg, \pTtruth, \pTtruth, $A_\mathrm{jet}$, and jet angularity. Within each panel, the evolution of \pTresid for events with no quenching to quenching with a \SI{8}{fm} brick of QGP.}
    \label{fig:resid_fm_prog_angularity}
\end{figure}

\begin{figure}[H]
    \centering
    \includegraphics[width=0.8\textwidth, trim=0.1cm 0.20cm 0.0cm 0.0cm, clip] {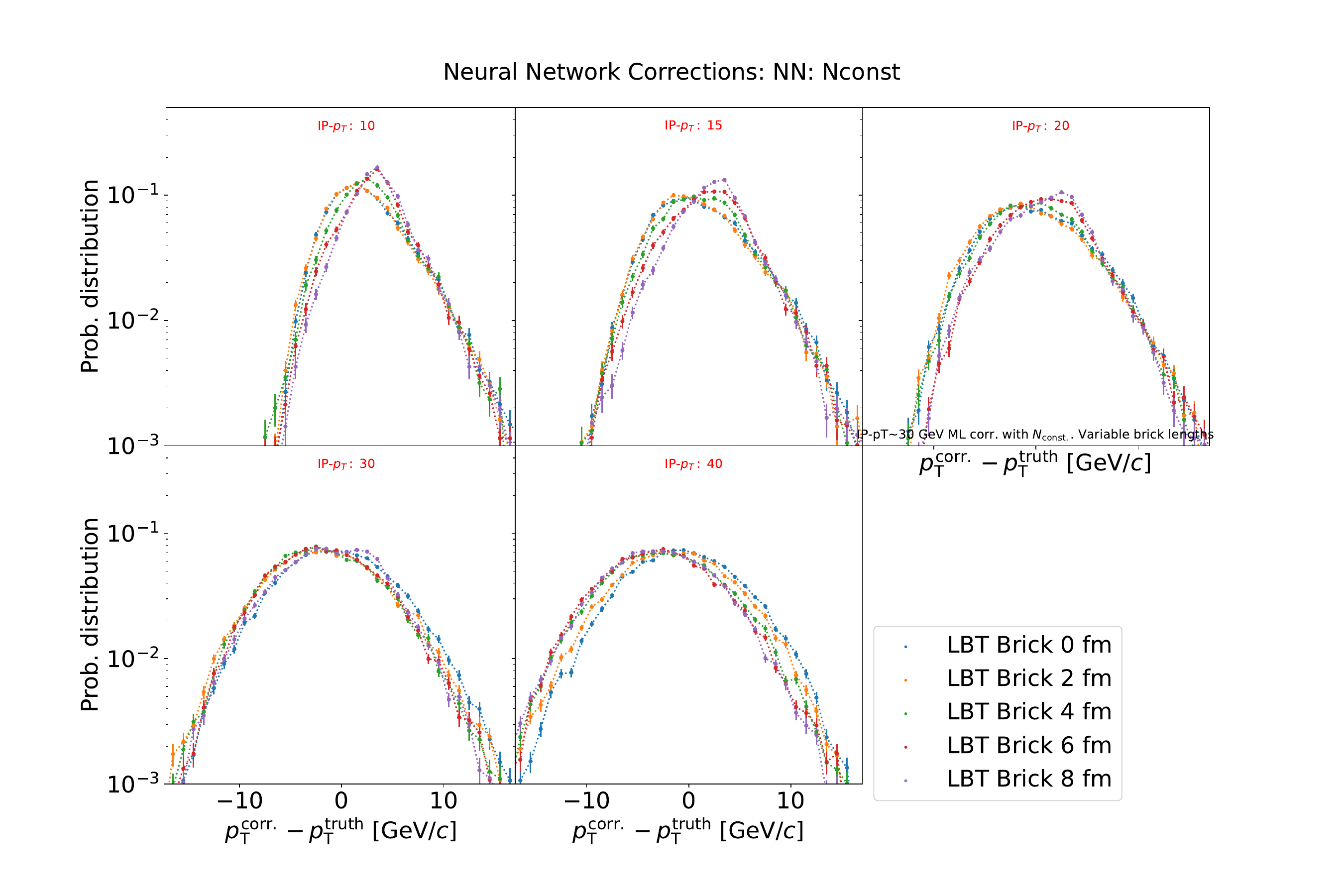}
    \caption{Distributions of $\pTresid\equiv\pTcorr-\pTtruth$ for distributions with correction made by a neural network trained on \rhobkg, \pTtruth, \pTtruth, $A_\mathrm{jet}$, and the number of jet constituents. Within each panel, the evolution of \pTresid for events with no quenching to quenching with a \SI{8}{fm} brick of QGP.}
    \label{fig:resid_fm_prog_nconsts}
\end{figure}

\begin{figure}[H]
    \centering
    \includegraphics[width=0.8\textwidth, trim=0.1cm 0.20cm 0.0cm 0.0cm, clip] {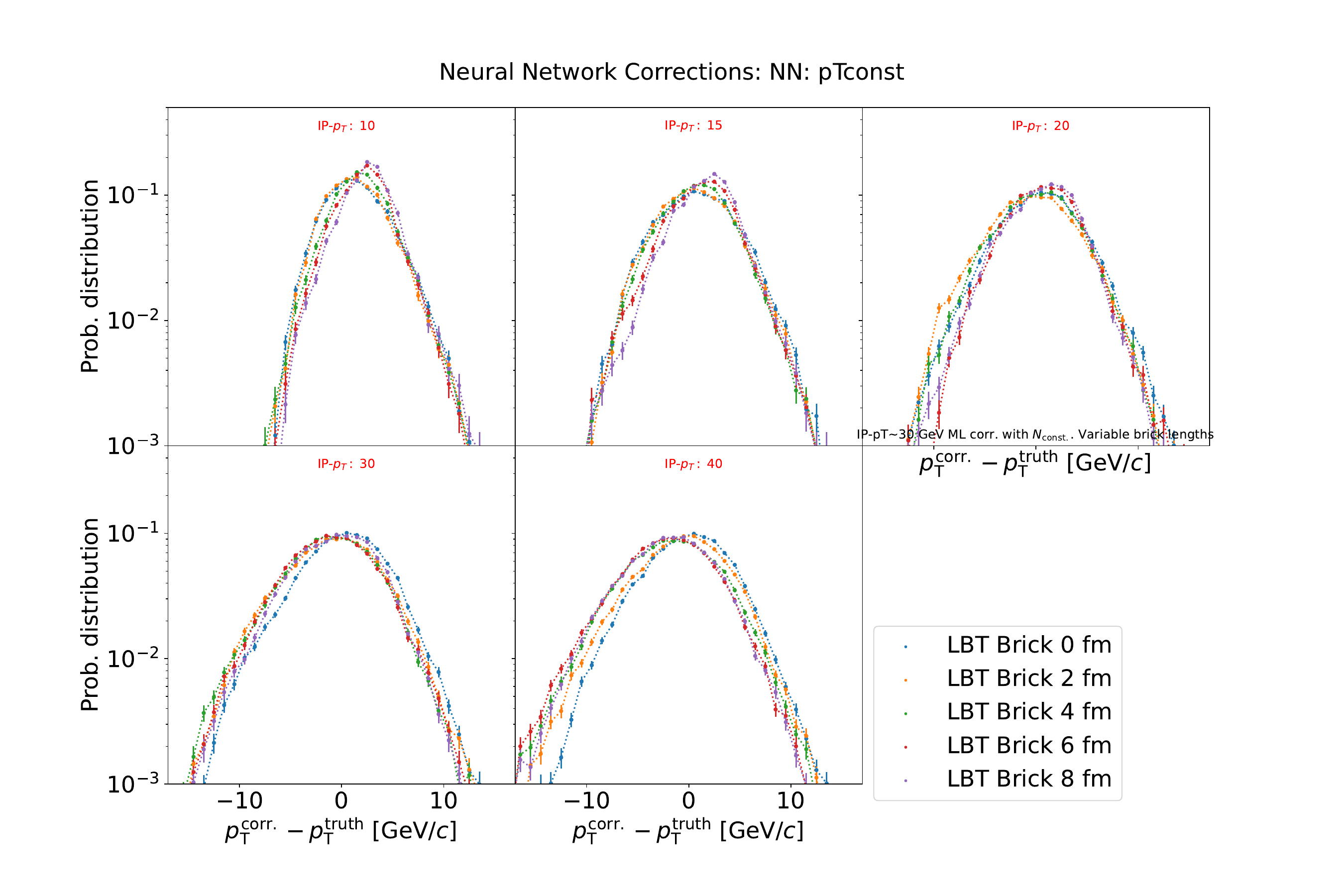}
    \caption{Distributions of $\pTresid\equiv\pTcorr-\pTtruth$ for distributions with correction made by a neural network trained on \rhobkg, \pTtruth, \pTtruth, $A_\mathrm{jet}$, the \pT of the leading 10 jet constituents. Within each panel, the evolution of \pTresid for events with no quenching to quenching with a \SI{8}{fm} brick of QGP.}
    \label{fig:resid_fm_prog_ptconst}
\end{figure}

\begin{figure}[H]
    \centering
    \includegraphics[width=0.8\textwidth, trim=0.1cm 0.20cm 0.0cm 0.0cm, clip] {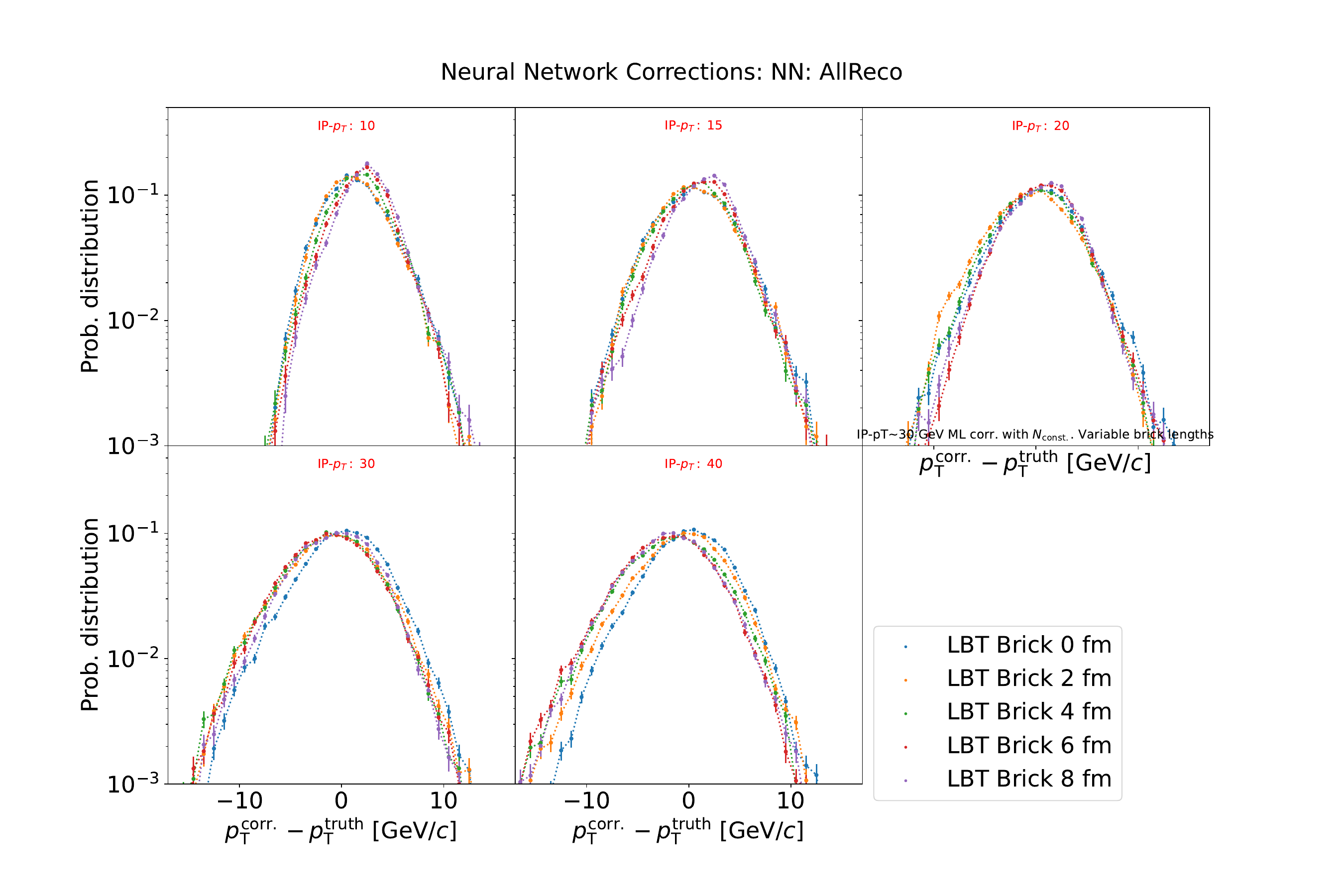}
    \caption{Distributions of $\pTresid\equiv\pTcorr-\pTtruth$ for distributions with correction made by a neural network trained on \rhobkg, \pTtruth, \pTtruth, $A_\mathrm{jet}$, jet angularity, the number of jet constituents, and the \pT of the leading 10 jet constituents. Within each panel, the evolution of \pTresid for events with no quenching to quenching with a \SI{8}{fm} brick of QGP.}
    \label{fig:resid_fm_prog_recoall}
\end{figure}



\begin{figure}[H]
    \centering
    \subfigure[Neural network trained with \rhobkg, $A_\mathrm{jet}$, \pTtruth, and \pTreco]{
        \includegraphics[width=0.45\textwidth, trim=0.4cm 0.80cm 0.6cm 2.0cm, clip] {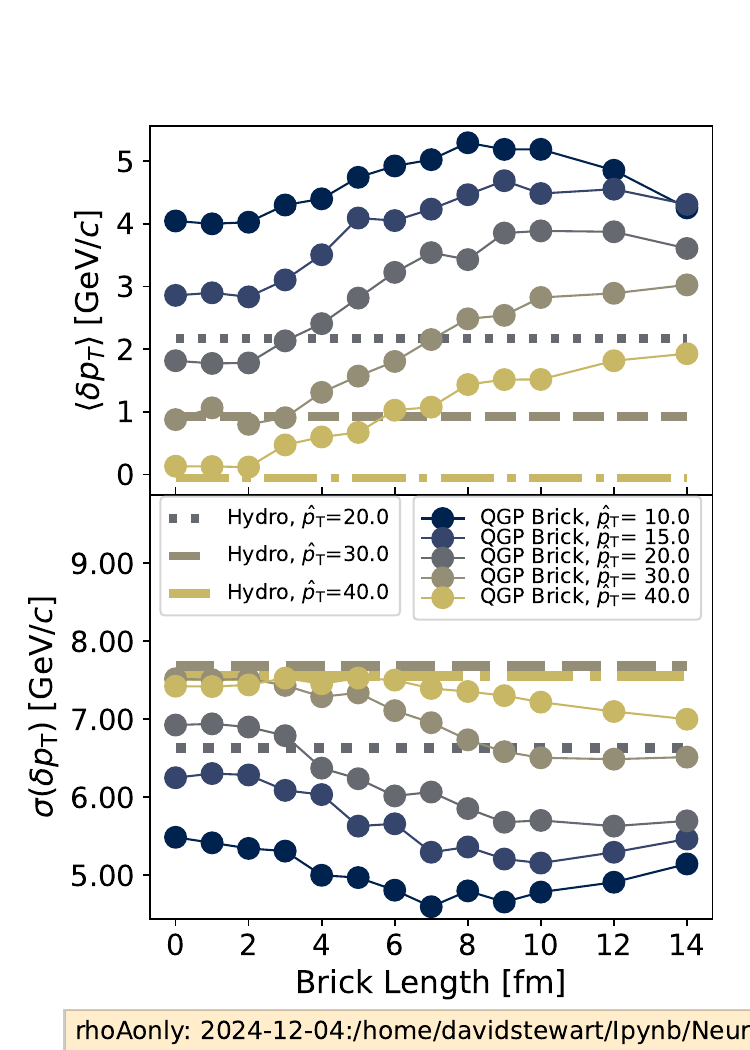}
        \label{fig:subfig1}
    }
    \hfill
    \subfigure[Neural network trained with \rhobkg, $A_\mathrm{jet}$, \pTtruth, \pTreco, and jet angularity]{
        \includegraphics[width=0.45\textwidth, trim=0.4cm 0.80cm 0.6cm 2.0cm, clip] {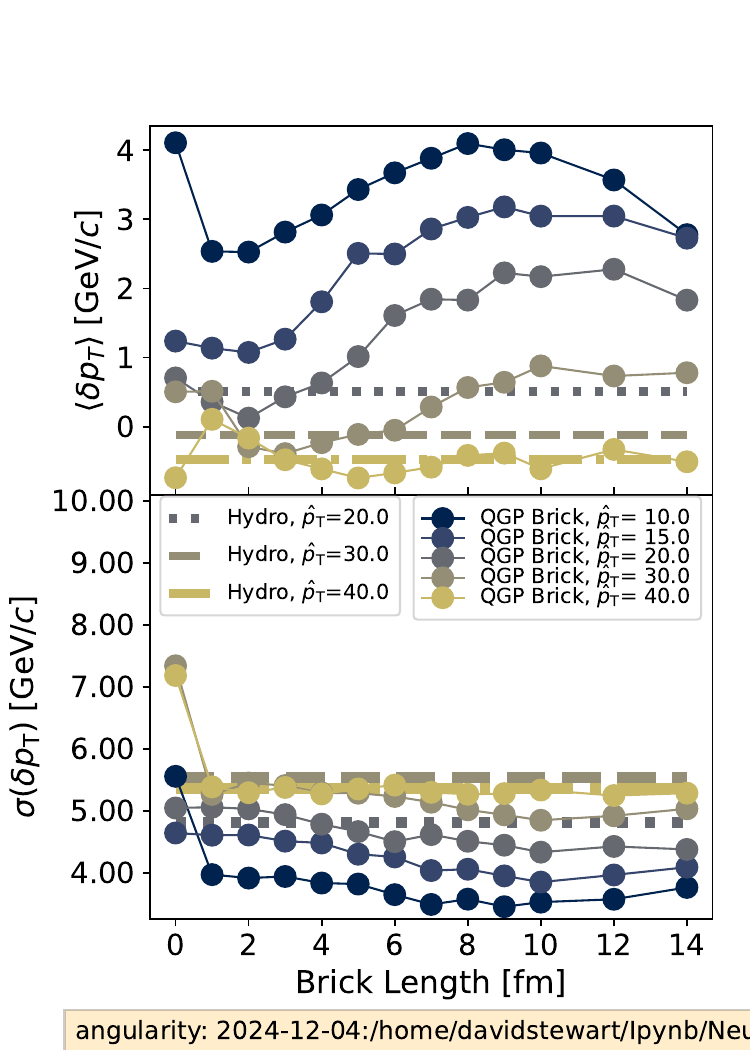}
        \label{fig:subfig2}
    }
    \caption{Mean and standard deviations of the $\pTresid\equiv\pTcorr-\pTtruth$ distributions for \pTcorr generated from a neural network trained on jet events with no quenching using all parameters listed in in Table~\ref{tab:NNinputs} (i.e. row (e)). The jets corrected are generated using hydrodynamic simulated QGP, as well as bricks of QGP, and (for ``Brick Length=0'') no quenching. The hydro data aren't associated with set brick lengths, and are displayed with horizontal lines at their $\langle\pTresid\rangle$ and $\sigma(\pTresid)$ values.}
    \label{fig:mean_std_prog_rhoA_and_angle}
\end{figure}

\begin{figure}[H]
    \centering
    \subfigure[Neural network trained with \rhobkg, $A_\mathrm{jet}$, \pTtruth, \pTreco, and number of jet constituents]{
        \includegraphics[width=0.45\textwidth, trim=0.4cm 0.80cm 0.6cm 2.0cm, clip] {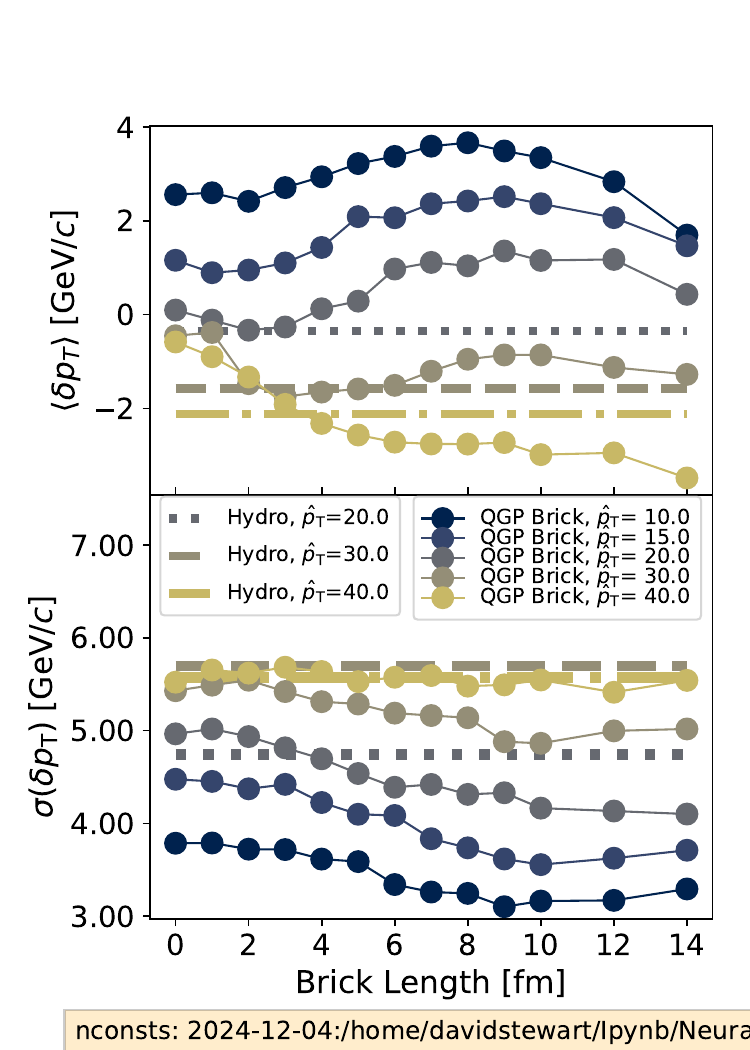}
        \label{fig:subfig3}
    }
    \hfill
    \subfigure[Neural network trained with \rhobkg, $A_\mathrm{jet}$, \pTtruth, \pTreco, and \pT of the jet's ten highest-\pT constituents]{
        \includegraphics[width=0.45\textwidth, trim=0.4cm 0.80cm 0.6cm 2.0cm, clip] {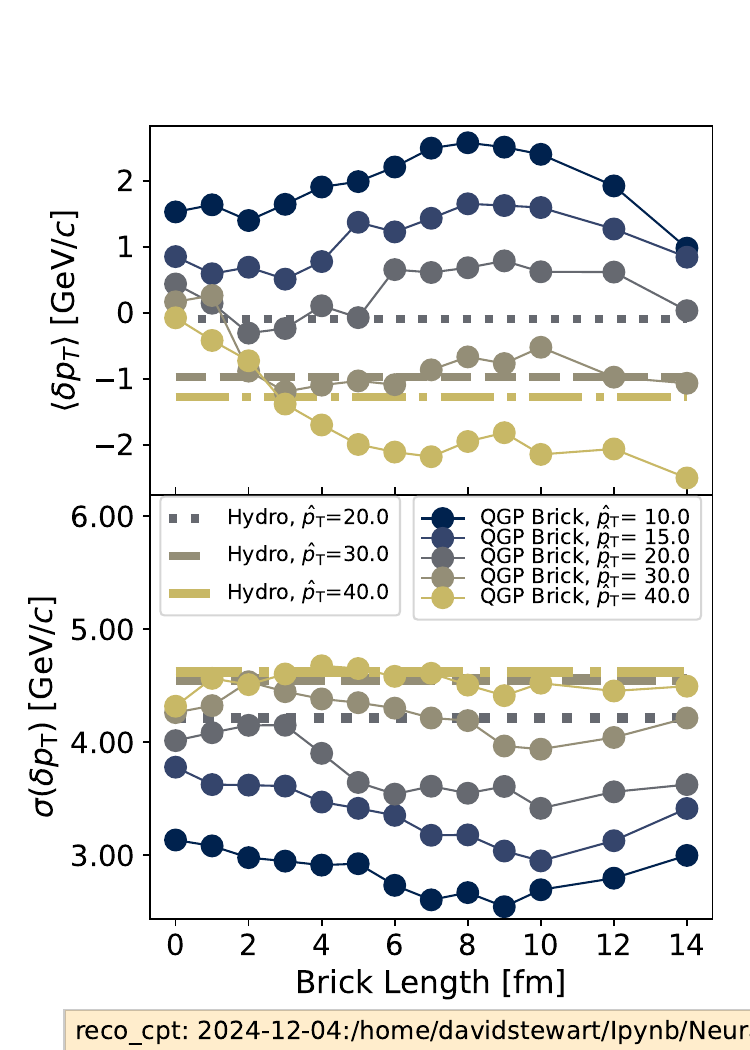}
        \label{fig:subfig4}
    }
    \caption{Mean and standard deviations of the $\pTresid\equiv\pTcorr-\pTtruth$ distributions for \pTcorr generated from a neural network trained on jet events with no quenching using all parameters listed in in Table~\ref{tab:NNinputs} (i.e. row (e)). The jets corrected are generated using hydrodynamic simulated QGP, as well as bricks of QGP, and (for ``Brick Length=0'') no quenching. The hydro data aren't associated with set brick lengths, and are displayed with horizontal lines at their $\langle\pTresid\rangle$ and $\sigma(\pTresid)$ values.}
    \label{fig:mean_std_prog_nconst_and_cpt}
\end{figure}

\begin{figure}[H]
    \centering
    \subfigure[Using neural network \NN{Ncons}, which is trained with \rhobkg, $A_\mathrm{jet}$, \pTtruth, \pTreco, and number of jet constituents.]{
        \includegraphics[width=0.45\textwidth, trim=1.4cm 0.80cm 1.5cm 0.0cm, clip] {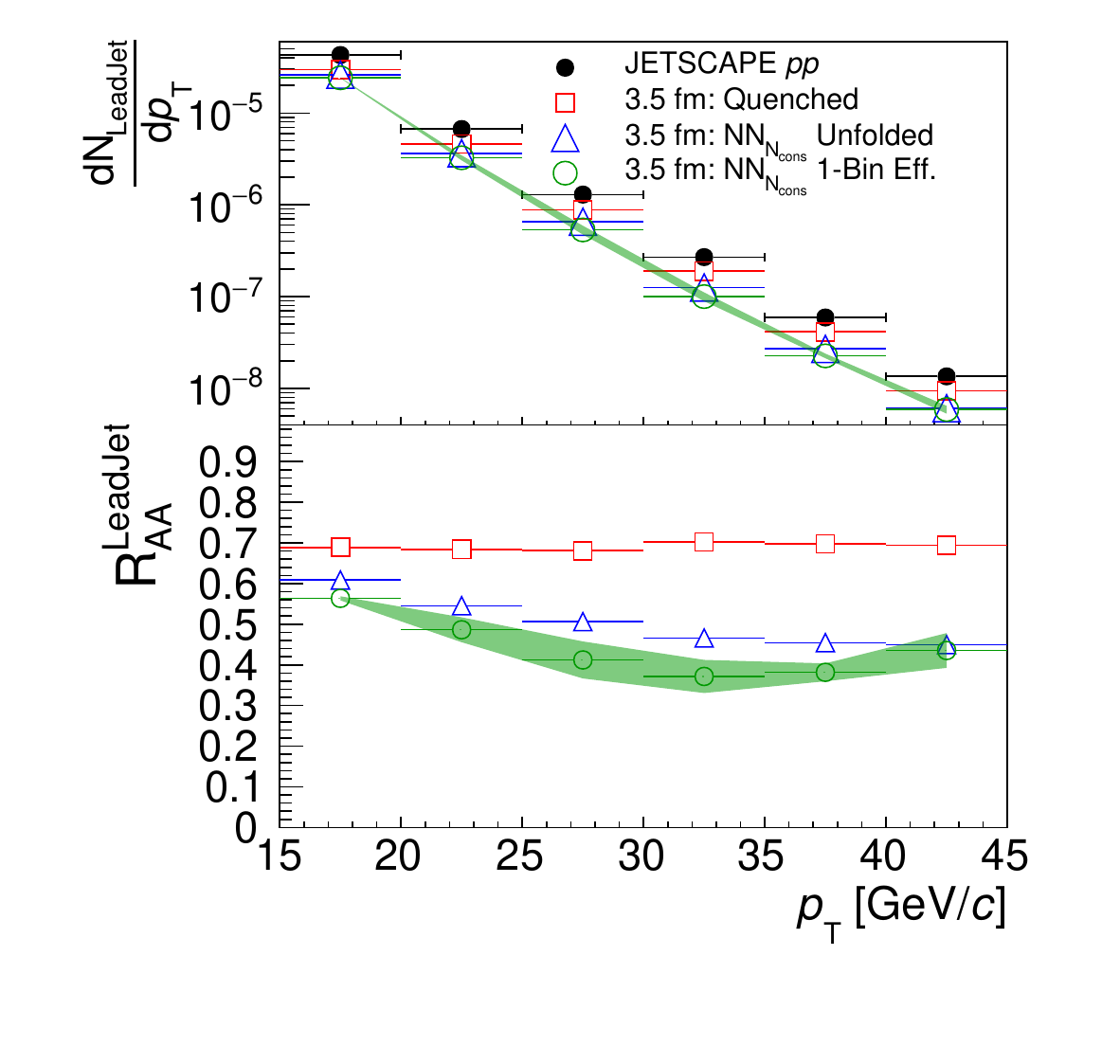}
        \label{fig:subfig1}
    }
    \hfill
    \subfigure[Using neural network \NN{Ang}, which is trained with \rhobkg, $A_\mathrm{jet}$, \pTtruth, \pTreco, and the jet angularity.]{
        \includegraphics[width=0.45\textwidth, trim=1.4cm 0.80cm 1.5cm 0.0cm, clip] {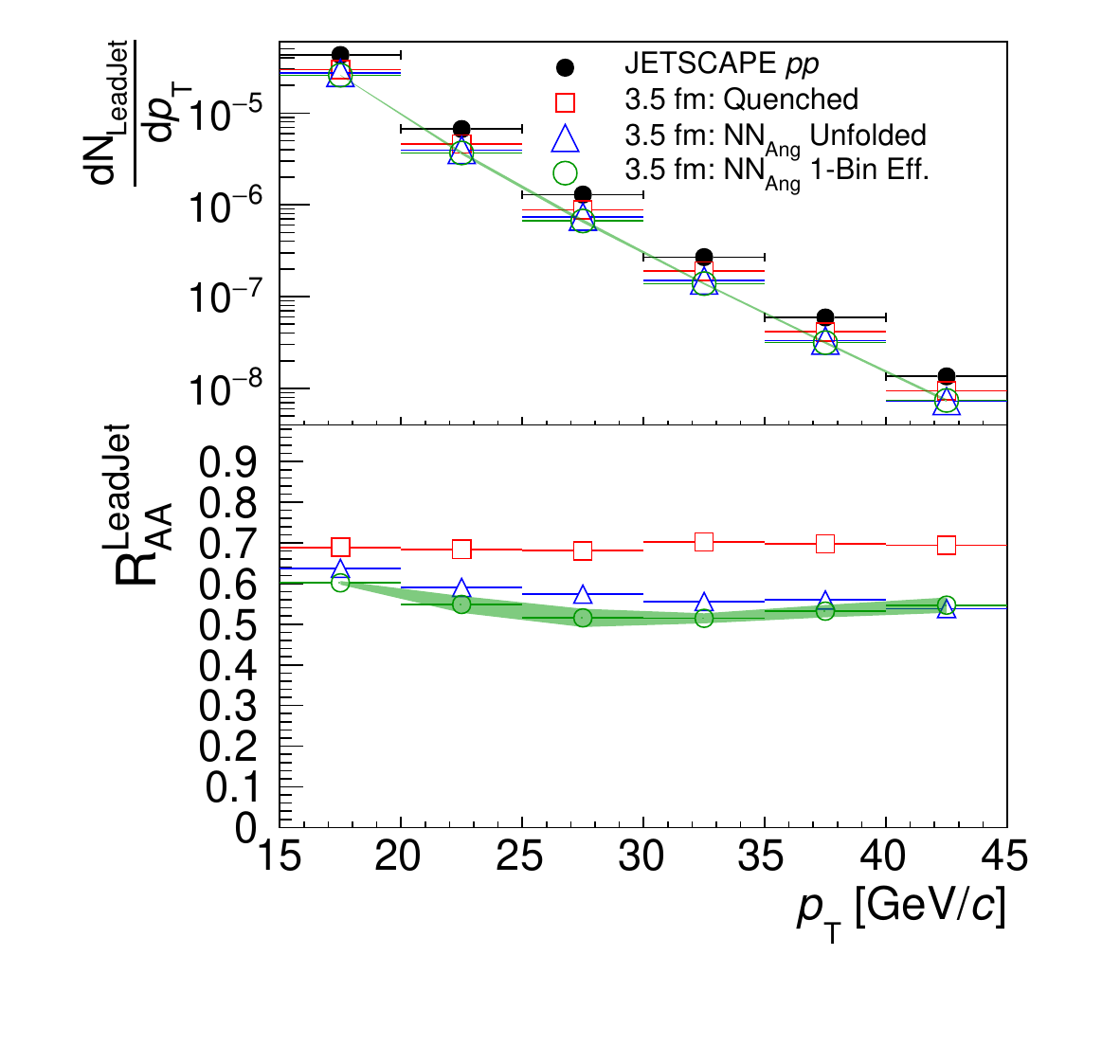}
        \label{fig:subfig2}
    }
    \caption{Jet spectra unquenched, quenched in \SI{3.5}{fm} of QGP, and the measured quenched jet spectra corrected for background using Neural network's with the parameters listed. (Training parameters are also listed in Table.~\ref{tab:NNinputs}.}
    \label{fig:RAA_like_Ang_Ncons}
\end{figure}

\begin{figure}[H]
    \centering
    \subfigure[Using neural network \NN{Ncons}, which is trained with \rhobkg, $A_\mathrm{jet}$, \pTtruth, \pTreco, and the \pT of highest 10 \pT jet constituents.]{
        \includegraphics[width=0.45\textwidth, trim=1.4cm 0.80cm 1.5cm 0.0cm, clip] {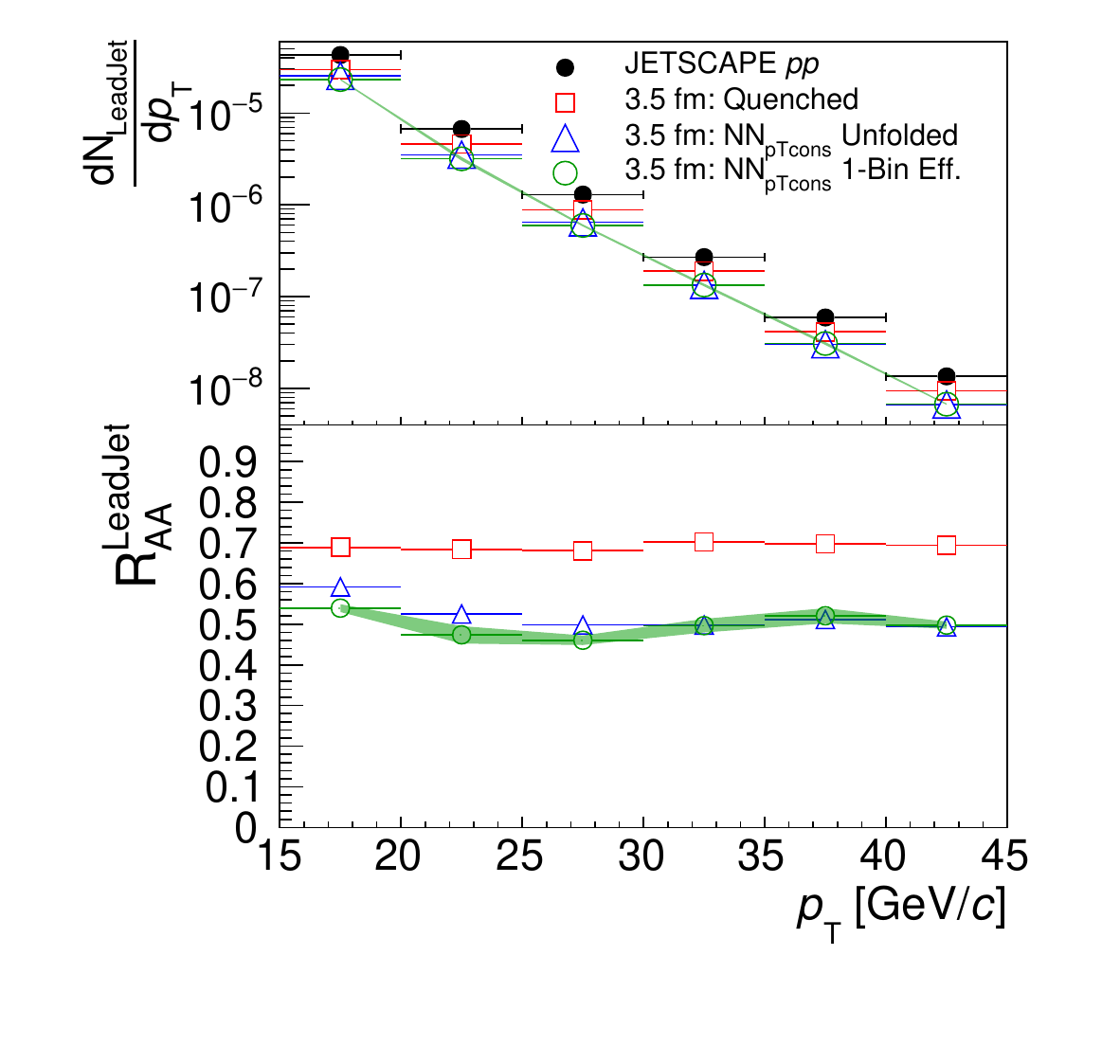}
        \label{fig:subfig3}
    }
    \hfill
    \subfigure[Using neural network \NN{Ang}, which is trained with \rhobkg, $A_\mathrm{jet}$, \pTtruth, \pTreco.]{
        \includegraphics[width=0.45\textwidth, trim=1.4cm 0.80cm 1.5cm 0.0cm, clip] {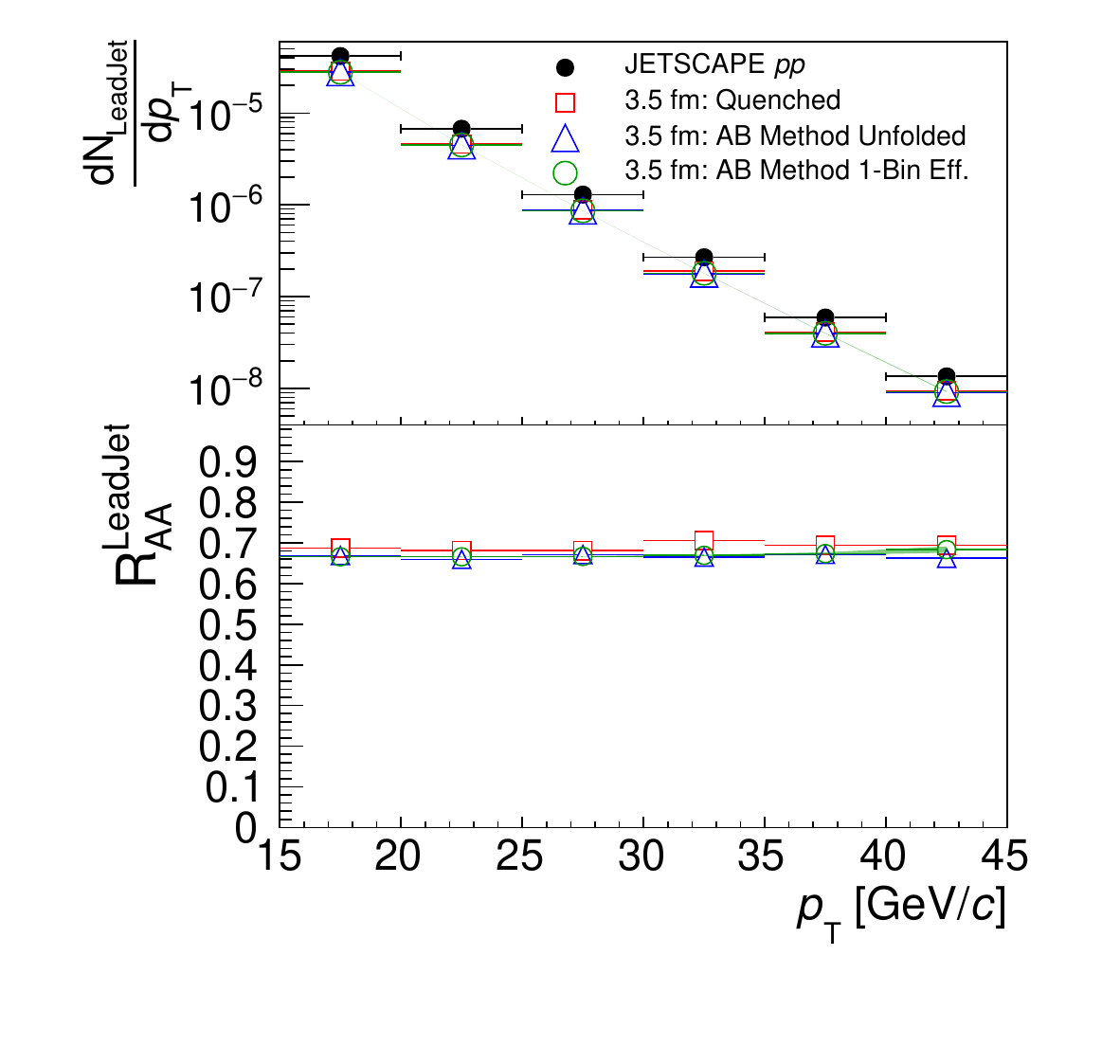}
        \label{fig:subfig4}
    }
    \caption{Jet spectra unquenched, quenched in \SI{3.5}{fm} of QGP, and the measured quenched jet spectra corrected for background using Neural network's with the parameters listed. (Training parameters are also listed in Table.~\ref{tab:NNinputs}.}
    \label{fig:RAA_like_cpt_AB}
\end{figure}

\begin{figure}[h]
    \centering
    \includegraphics[width=0.45\textwidth, trim=3.3cm 0.80cm 1.9cm 0.7cm, clip] {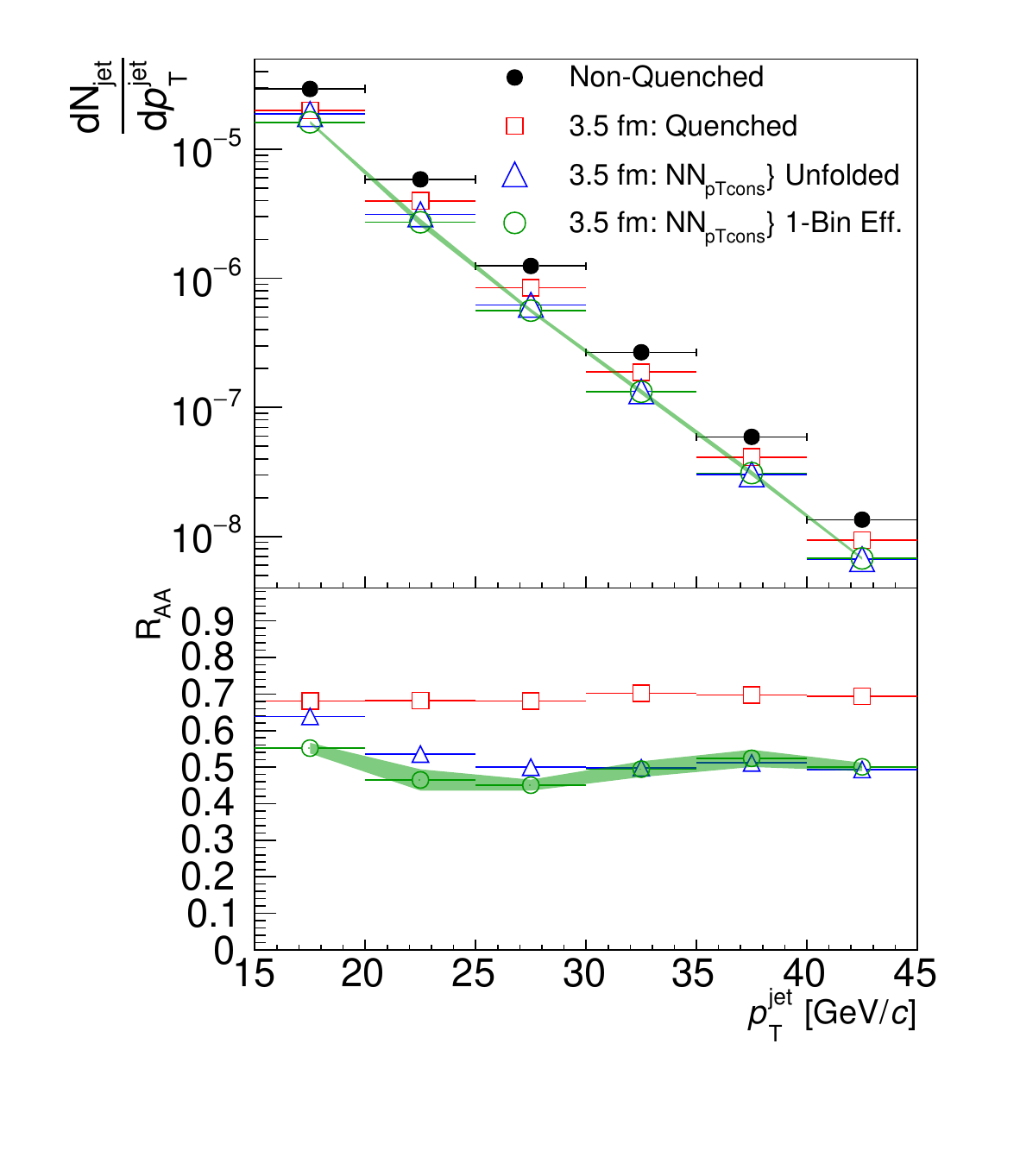}
    \caption{Jet spectra unquenched, quenched in \SI{3.5}{fm} of QGP, and the measured quenched jet spectra corrected for background using neural network \NN{pTcons}, which is trained with \rhobkg, $A_\mathrm{jet}$, \pTtruth, \pTreco, and \pT of the jet's ten highest-\pT constituents.}
    \label{fig:RAA_like_cpt}
\end{figure}


\section{Hardware and Software}\label{app:software}

\subsection{Hardware}

All code was run on a single machine equipped with an AMD Ryzen Threadripper 3960X Processor, two NVIDIA GeForce RTX 3090 GPUs, and 128 GB of DDR4 ram.

\subsection{Software}

The following software process was used. Input files, scripts, and codes, are archived at \texttt{github.site}.

\begin{itemize}
    \item JETSCAPE 3.6.4 \cite{Putschke:2019yrg} was pulled from online at \texttt{https://github.com/JETSCAPE/JETSCAPE}, compiled locally, and run with \texttt{XML} files.
    \item JETSCAPE's output \texttt{.dat.gz} files were converted into \texttt{ROOT} \cite{Brun:2000es} files via a Python script.
    \item A locally compiled C++ code, using \texttt{ROOT} 6.28/10 \cite{Brun:2000es} and \texttt{FastJet} 3.4.2 \cite{Cacciari:2011ma} libraries, was used to cluster and match jets, calculated \rhobkg, etc... The output files were \texttt{ROOT} files.
    \item Python scripts and Jupyter Notebooks were used to process the output files and run the machine learning. The principle Python libraries used are:
    \begin {itemize}
        \item Python 3.10.12
        \item Pandas 2.2.1
        \item Scikit-learn 0.23.2 \cite{Pedregosa:2011ork}
        \item NumPy 1.26.4
        \item Pickle 4.0
        \item json 2.0.9
        \item TensorFlow 2.13.1 \cite{Abadi:2016kic}
        \item PyArrow 15.0.1
    \end{itemize}
\end{itemize}

\newpage
\bibliographystyle{apsrev4-2}
\bibliography{main}
\end{document}